\documentclass[aps,prl,nobibnotes,twocolumn,superscriptaddress,citeautoscript,showkeys]{revtex4-2}

\pdfoutput=1
\usepackage{graphicx}
\usepackage{amsmath, amsfonts}
\usepackage{amssymb}
\usepackage[colorlinks=true,linkcolor=blue,citecolor=blue]{hyperref}
\usepackage[utf8]{inputenc}
\usepackage{float}
\usepackage{subcaption}
\usepackage{ragged2e}

\DeclareCaptionJustification{justified}{\justifying}
\setcitestyle{super}

\begin{document}

\title{VO$_2$ oscillator circuits optimized for ultrafast, 100 MHz-range operation}

\author{Zsigmond Pollner}\altaffiliation{Contributed equally to this work}
\affiliation{Department of Physics, Institute of Physics, Budapest University of Technology and Economics, M\H{u}egyetem rkp. 3., H-1111 Budapest, Hungary.}
\affiliation{HUN-REN–BME Condensed Matter Research Group,
Budapest University of Technology and Economics,
Műegyetem rkp. 3., H-1111 Budapest, Hungary.}

\author{Tímea Nóra Török}\altaffiliation{Contributed equally to this work}
\affiliation{Department of Physics, Institute of Physics, Budapest University of Technology and Economics, M\H{u}egyetem rkp. 3., H-1111 Budapest, Hungary.}
\affiliation{Institute of Technical Physics and Materials Science, Centre for Energy Research, Konkoly-Thege M. \'{u}t 29-33, 1121 Budapest, Hungary.}

\author{László Pósa}
\affiliation{Department of Physics, Institute of Physics, Budapest University of Technology and Economics, M\H{u}egyetem rkp. 3., H-1111 Budapest, Hungary.}
\affiliation{Institute of Technical Physics and Materials Science, Centre for Energy Research, Konkoly-Thege M. \'{u}t 29-33, 1121 Budapest, Hungary.}

\author{{Miklós Csontos}}
\affiliation{Institute of Electromagnetic Fields, ETH Zurich, Gloriastrasse 35, 8092 Zurich, Switzerland.}

\author{{Sebastian Werner Schmid}}
\affiliation{Department of Physics, Institute of Physics, Budapest University of Technology and Economics, M\H{u}egyetem rkp. 3., H-1111 Budapest, Hungary.}
\affiliation{Experimental Physics V, Center for Electronic Correlations and Magnetism, University of Augsburg, 86135 Augsburg, Germany.}

\author{Zoltán Balogh}
\affiliation{Department of Physics, Institute of Physics, Budapest University of Technology and Economics, M\H{u}egyetem rkp. 3., H-1111 Budapest, Hungary.}
\affiliation{HUN-REN–BME Condensed Matter Research Group,
Budapest University of Technology and Economics,
Műegyetem rkp. 3., H-1111 Budapest, Hungary.}

\author{András Bükkfejes}
\affiliation{Emerson - Test and Measurement (NI), Hungária körút 30/A, H-1087 Budapest, Hungary.}

\author{Heungsoo Kim}
\affiliation{Naval Research Laboratory, 4555 Overlook Ave, Washington, DC 20375, USA.}

\author{Alberto Piqué}
\affiliation{Naval Research Laboratory, 4555 Overlook Ave, Washington, DC 20375, USA.}

\author{{Jeurg Leuthold}}
\affiliation{Institute of Electromagnetic Fields, ETH Zurich, Gloriastrasse 35, 8092 Zurich, Switzerland.}

\author{János Volk}
\affiliation{Institute of Technical Physics and Materials Science, Centre for Energy Research, Konkoly-Thege M. \'{u}t 29-33, 1121 Budapest, Hungary.}

\author{András Halbritter}
\email{halbritter.andras@ttk.bme.hu}
\affiliation{Department of Physics, Institute of Physics, Budapest University of Technology and Economics, M\H{u}egyetem rkp. 3., H-1111 Budapest, Hungary.}
\affiliation{HUN-REN–BME Condensed Matter Research Group,
Budapest University of Technology and Economics,
Műegyetem rkp. 3., H-1111 Budapest, Hungary.}

\begin{abstract}
Oscillating neural networks are promising candidates for a new computational paradigm, where complex optimization problems are solved by physics itself through the synchronization of coupled oscillating circuits. Nanoscale VO$_2$ Mott memristors are particularly promising building blocks for such oscillating neural networks. Until now, however, not only the maximum frequency of VO$_2$ oscillating neural networks, but also the maximum frequency of individual VO$_2$ oscillators has been severely limited, which has restricted their efficient and energy-saving use. In this paper, we show how the oscillating frequency can be increased by more than an order of magnitude into the 100 MHz range by optimizing the sample layout and circuit layout. In addition, the physical limiting factors of the oscillation frequencies are studied by investigating the switching dynamics. To this end, we investigate how much the set and reset times slow down under oscillator conditions compared to the fastest switching achieved with single dedicated pulses. These results pave the way towards the realization of ultra-fast and energy-efficient VO$_2$-based oscillating neural networks.
\end{abstract}


\date{\today}
\maketitle

\section{Introduction}
The possibility of performing computational operations by the synchronization of coupled oscillators was originally raised by John von Neumann,\cite{Neumann1957,Wigington1959} and has since been demonstrated in various physical systems \cite{Todri-Sanial2024_Computing_with_oscillators_from_theoretical_to_appl, Mallick2020_coupledOscillatorsUsing65nmCmos, PhysRevResearch_6_023162_mechanicalMicroResonatorsCoupling, 9377463_RingOscillators_CombinatorialProblemCalculation}. This elegant idea is based on encoding the computational or optimization problem into the couplings between oscillators, and then the solution is provided by physics itself through the synchronization of the oscillators into different phase shifts.\cite{Albertsson2023_ringOscillatorsNetworkSomeTheory, Wang2021_isingMachines_problemSolving_moreTheory, Raychowdhury_2019_8565896_ONN_network_VO2is_CsatDinam} This scheme can be implemented with conventional semiconductor circuits such as ring oscillators,\cite{Datta2017} while relaxation oscillators made from novel memristive nano-devices enable even more compact and energy-efficient oscillator networks.\cite{CsabaGy2020} The most promising memristive oscillators rely on the nanoscale voltage-induced insulator-to-metal transition in Mott-type NbO$_2$ or VO$_2$ memristors.\cite{Chen2025_NbO2BasedOscillators_ThermalDesignAndNeuronNetwork, Lu2024_NbOx_oscillators_14MHzreachedRoughly_NoNetwork, Li2015_SomeNbOx_oscillators, 11MHz_Bohaichuk_2019_FastSpikingOfVO2CarbonNanotubeOscillator, Yi2018_Biological_plausibility_and_stochasticity_in_scalable_VO2_23spikingBewhaviours, Shukla2014_correlated_phases_Vo2Oscillators} The latter material system, which is the main focus of the present paper as well, has been successfully used to assemble oscillating neural networks (ONNs), which could solve problems like map coloring and maximum cut,\cite{Maher2024,Dutta2021_isingHamiltonianSolver_VO2_based_ONN_optimisation,Mohseni2022} gesture recognition,\cite{Yang2024, Yuan2022_calibratableSensoryNeuron_gestureRecognition} or feature extraction in convolutional neural networks.\cite{Corti2021, Corti2020_TimeDelayEncodedImageRecog_VO2_osc_8986576} 

\begin{figure}[t!]
	\centering
 \includegraphics[width=0.48\textwidth]{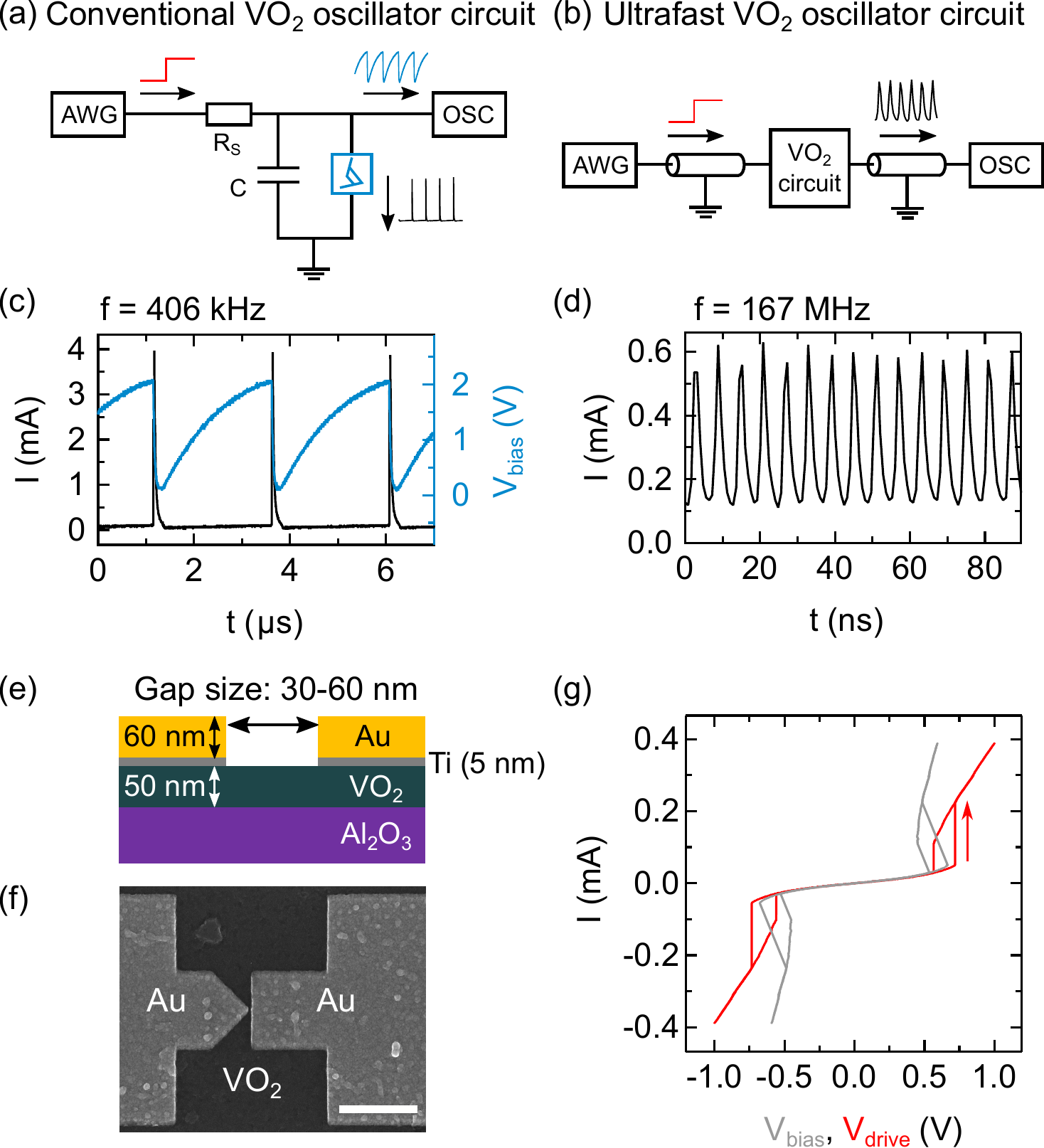}
	\caption{\it \textbf{Basics of our low-frequency and high-frequency VO$_2$ oscillators.} (a) Conventional low frequency oscillator circuit with a VO$_2$ memristor (blue box), and a series resistor and parallel capacitor. An arbitrary waveform generator (AWG) is used to switch the driving voltage from zero to the desired DC value (red step function), which yields an oscillating current (black) and voltage (blue) signal. The latter is measured by an oscilloscope (OSC).   (b) Scheme of the high-frequency transmission line oscillator arrangement. (c) Low-frequency oscillating waveforms produced by a conventional oscillator circuit (a). The measured oscillating current and voltage signals on the memristor are respectively shown by the black and blue lines. (d) Demonstration of the current signal of our highest frequency oscillation ($f=167\,$MHz) (e) Schematic of the memristor device highlighting the small device size and the layer structure (see the Methods section for more details). (f) Scanning electron micrograph of a representative VO$_2$ memristor, scale bar: 1~$\mu$m. The V-shaped sample layout focuses the switching to an ultrasmall spot. (g) Low-frequency I(V) graph of a VO$_2$ device as a function of driving voltage (red) and the bias voltage (grey) measured on the memristor.}
	\label{fig1}
\end{figure}

The building blocks of VO$_2$ ONNs are simple oscillator circuits, like the scheme illustrated in Fig.~\ref{fig1}a. The oscillation is granted by the hysteretic switching of the VO$_2$ memristor (blue square) and the resistor in series ($R_S$), while the oscillation frequency is tunable by the parallel capacitor $C$. This circuit produces the typical oscillating voltage (blue) or current (black) signal demonstrated in Fig.~\ref{fig1}c, once the input signal is switched from zero to a DC level (see red input signal in Fig.~\ref{fig1}a). 

The performance of ONNs is directly influenced by the frequency of the oscillations; higher oscillation frequencies enable faster and more energy-efficient computations, thereby reducing the overall computing time and energy consumption. \cite{Maher2024,10026654_VO2_ONNlinedetection_SmallVO2FasterOscSimulation} According to the state of the art, the oscillation frequency of single VO$_2$ oscillators have not yet exceeded the maximum of $9\,$MHz,\cite{10.1063/1.4922122_9MHz_cikk} while VO$_2$ oscillating neural networks were usually operated at even lower frequencies, such as $2\,$kHz\cite{Maher2024} or $3\,$MHz\cite{Li2024_3MHz_ThermallyCopledVO2Memristors}. In TaO$\rm_X$ oscillators a record frequency of $250\,$MHz was achieved,\cite{7283590_TaO_X_250MHz_Oscillator_2015} however, in that case the frequency was boosted by an active transistor in the circuit.

In this paper, we investigate the possibility of increasing the operating frequency in circuits that contain only the passive series resistor and parallel capacitance in addition to the Mott memristor. In particular, we demonstrate significantly faster oscillations in VO$_2$ devices compared to the previous studies,\cite{10.1063/1.4922122_9MHz_cikk, 11MHz_Bohaichuk_2019_FastSpikingOfVO2CarbonNanotubeOscillator} as demonstrated by our record $167\,$MHz oscillation frequency in Fig.~\ref{fig1}d. This achievement relies on three optimization aspects. (i) For the oscillators we use VO$_2$ devices, in which the operation is focused to an ultrasmall, few tens of nanometers wide active region,\cite{doi:10.1021/acsanm.3c00150_Laci_Interplay_Cikk, 10026654_VO2_ONNlinedetection_SmallVO2FasterOscSimulation} while the stray capacitance is minimized. This  allows us to exceed theoretical frequency limits established for so-called crossbar VO$_2$ devices\cite{Carapezzi_2023_HowFastVO2CanOscillate}. Similar, strongly confined operation region devices were applied in our previous study to demonstrate the fastest electrically induced VO$_2$ switching so far.\cite{doi:10.1021/acsnano.4c03840_Sebastian_PicojouleSwitching} (ii) The oscillator circuit layout is optimized for high-frequency, transmission-line geometry (see illustration in Fig.~\ref{fig1}b). (iii) The physical limitations of frequency boosting are demonstrated through examining the internal physical relaxation time of the VO$_2$ memristor.

These steps are presented as follows. First, the fabrication and characterization of the optimized VO$_2$ devices is summarized. Afterwards, the geometric frequency limitations and the optimized circuit layout are motivated by transmission-line geometry LTspice simulations. Next, we summarize our experimental results on the ultrafast oscillating circuits. Finally, the physical limitations of the oscillation frequency are analyzed through pulsed relaxation-time experiments. 


\section{Results and Discussion}

\subsection{Sample characterisation and low-frequency measurements}

The schematic vertical cross-section of our planar, $\approx$30~nm gap size devices is shown in {Fig.~\ref{fig1}e}. For the high-frequency experiments, the use of a doped Si substrate would be disadvantageous due to the high stray capacitance towards the substrate, so a well insulating Al$_2$O$_3$ substrate was applied. The gold top electrodes were patterned on the epitaxial VO$_2$ layer by standard electron beam lithography (see the Methods section for fabrication details). A key feature of the structure is the asymmetric lateral geometry (Fig.~\ref{fig1}f): a rectangular electrode on one side faces a V-shaped electrode on the other side. This geometry enables the production of ultrasmall ($\approx 30\,$nm) gap sizes and the confinement of the active region in an ultranarrow, nanoscale spot.\cite{doi:10.1021/acsanm.3c00150_Laci_Interplay_Cikk} 
This confined geometry is considered crucial for the ultrafast operation, as well as for achieving sufficiently low switching threshold voltages and low switching energies at room temperature.\cite{doi:10.1021/acsnano.4c03840_Sebastian_PicojouleSwitching} 


The electrical switching is represented by the $I(V)$ curves of our devices in {Fig.~\ref{fig1}g} where the red trace shows the measured $I$ current as a function of the \emph{drive voltage} ($V_{\rm drive}$) applied to the VO$_2$ memristor and the $R_{\rm S}=380 \, \Omega$ resistor in series, whereas the grey graph is the function of the $V_{\rm bias}=V_{\rm drive}-I\cdot R_{\rm S}$ voltage drop on the memristor.

Applying an even larger resistance in series ($R_{\rm S} > 4500 \, \Omega$) a self-oscillation of the VO$_2$ memristor can be induced by a constant driving voltage. In this case the $I=(V_{\rm drive}-V_{\rm bias})/R_{\rm S}$ load line does not cross the $I(V_{\rm bias})$ curve at stable states, i.e. the system started in the OFF state switches ON before reaching the stable high resistance state, but then the bias voltage is released due to the voltage division by the series resistor and the system switches back to the OFF state before reaching the stable low resistance state, and the whole process is periodically repeated.
The frequency of the self-oscillation is set by a parallel capacitor $C$, as shown in the circuit diagram of Fig.~\ref{fig1}a, such that the dynamics of the set (high resistance to low resistance, or OFF to ON) or reset (low resistance to high resistance, or ON to OFF) transition is governed by the RC time-constant of the capacitor and the OFF or ON state resistances of the VO$_2$ memristor. Applying a step function from an arbitrary waveform generator (AWG) on this circuit (red line in Fig.~\ref{fig1}a), i.e. increasing the drive voltage from zero to a sufficient DC value, yields an oscillating operation at the output (blue illustration Fig.~\ref{fig1}a), which can be measured by an oscilloscope (OSC).  The measured periodic oscillation of the voltage (blue) and current (black) signal is exemplified in Fig.~\ref{fig1}c. 

\begin{figure}[t!]
	\centering
 \includegraphics[width=0.48\textwidth]{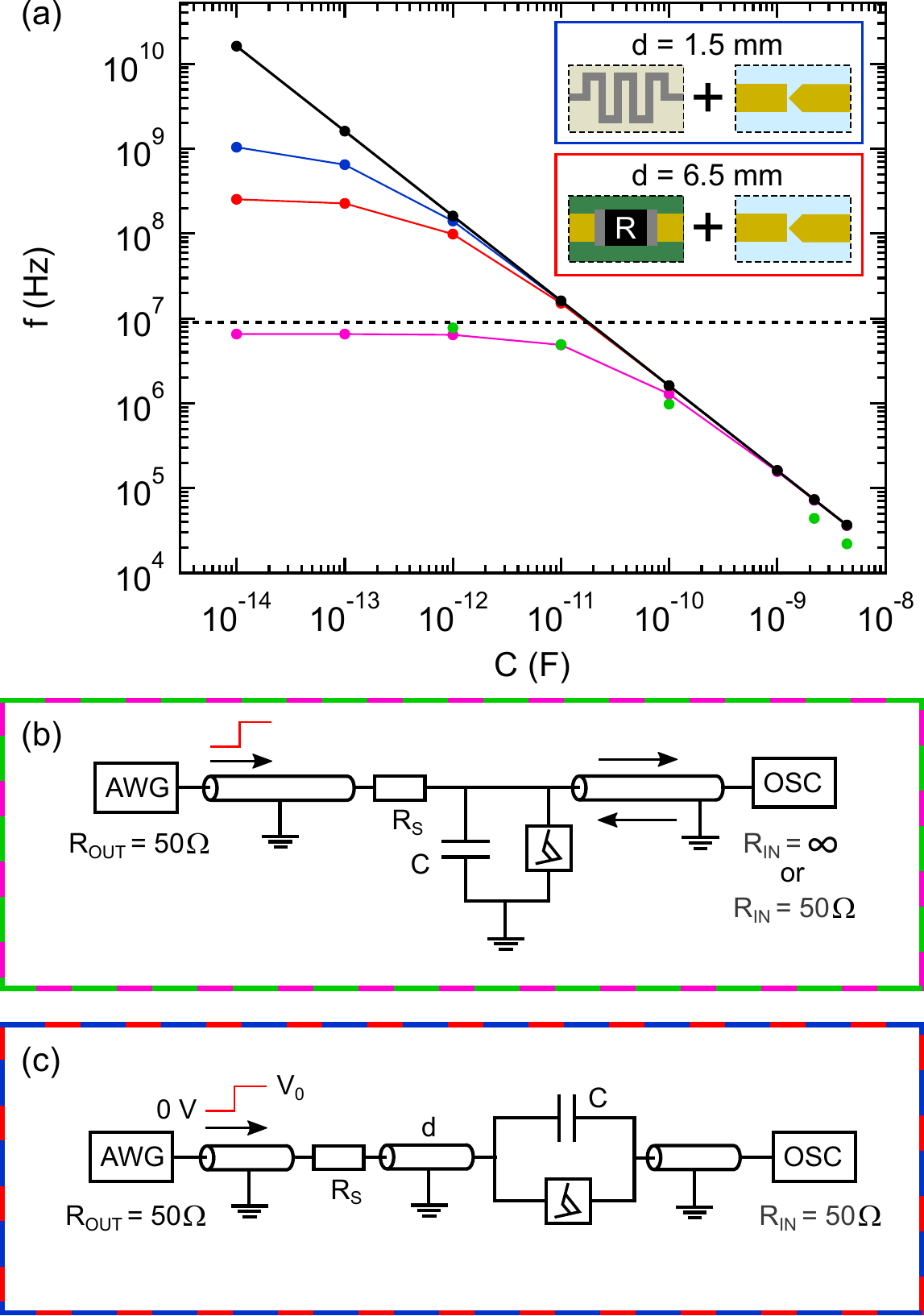}
	\caption{\it \textbf{Simulation of the oscillation frequencies using various circuits arrangements.} 
    (a) Oscillation frequencies as the function of the parallel $C$ capacitance in comparison to the highest $9\,$MHz oscillation frequency achieved so far (black dashed line).\cite{10.1063/1.4922122_9MHz_cikk} The black curve illustrates the simulated oscillation frequencies for a conventional VO$_2$ oscillator circuit (see Fig.\ \ref{fig1}a) such that the finite signal propagation speed in the cables is not taken into account at all. The same circuit arrangement yields the pink curve once the cable between the circuit and the high-impedance OSC is treated as a transmission line (see panel (b)) with $0.25\,$m length and  $v=2\cdot 10^8\,$m/s propagation speed. \cite{7543950_RG58Cable_propagationSpeed}  Note the low-pass filter nature of the oscilloscope-cable pair. These simulated frequency values are consistent with our experimental results (green dots) measured in the arrangement of panel (b). The red and blue curves correspond to the in-line oscillator arrangement of panel (c). This arrangement is optimized for high-frequency operation such that only the finite memristor-to-resistor distance $d$ is the limiting factor for the maximum achievable frequencies. This distance is respectively set to $d=6.5\,$mm and $d=1.5\,$mm for the red and blue curves. In the simulations experimentally reasonable circuit parameters of $R_S=22\,\mathrm{k}\Omega$, $R_\mathrm{M,OFF}=28\,\mathrm{k}\Omega$, $R_\mathrm{M,ON}=180\,\Omega$, $V_0=5\,$V, $V_\mathrm{set}=2.4\,$V and $V_\mathrm{reset}=0.45\,$V are applied. The red- and blue-framed insets illustrate possible experimental realizations of different distances: $d=6.5\,$mm is easily realized on a printed circuit board with a surface-mounted resistor, while $d=1.5\,$mm, or even smaller sub-mm distances already prefer an integrated on-chip resistor, like a meander-shaped platinum wire. (b) Conventional oscillator circuit, but the cables to the AWG and OSC are considered as transmission lines. (c) In-line oscillator circuit arrangement optimized for high-frequency operation, where every cable is modeled as a transmission line.}
	\label{fig2}
\end{figure}

In such a conventional VO$_2$ oscillator circuit (Fig.~\ref{fig1}a) the oscillation frequency is expected to be an inverse function of the $C$ parallel capacitance, as shown by the results of simple circuit simulations (Fig.~\ref{fig2}a black curve). In these simulations the VO$_2$ memristor is treated as a simple hysteretic switch with $V_\mathrm{set}$ and $V_\mathrm{reset}$ set and reset voltages and $R_\mathrm{M,OFF}$ and $R_\mathrm{M,ON}$ resistances in the high resistance OFF state and the low resistance ON state of the memristor (see the Methods section for details of the simulations). In practice, the parallel capacitance is limited by the stray capacitance of the device, which can be reduced to the level of a few femtofarads by careful sample design.\cite{doi:10.1021/acsnano.4c03840_Sebastian_PicojouleSwitching} According to our simple simulations,  a correspondingly chosen $\approx 10^{-14}\,$F minimal parallel capacitance would yield $\approx 10\,$GHz oscillation frequencies, while a conventionally achievable $\approx 10^{-12}\,$F parallel capacitance would result in $\approx 100\,$MHz oscillations (see black line in Fig.~\ref{fig2}a). These oscillation frequencies would highly exceed the current world record of $9\,$MHz (horizontal dashed line in Fig.~\ref{fig2}a) demonstrated in the fastest VO$_2$ oscillator circuits so far.\cite{10.1063/1.4922122_9MHz_cikk} Our experiments with the oscillator circuit of Fig.~\ref{fig1}a result in the green data points in Fig.~\ref{fig2}a, which also show a significantly slower experimental oscillation in the low capacitance range than expected (i.e. compared to the black curve). This deviation is attributed to the fact, that the circuit of Fig.~\ref{fig1}a does not consider the finite signal propagation speed in the lines. The observed slow-down is explained by our generalized LTspice simulations (pink curve in Fig.~\ref{fig2}a), where the cables connecting the oscillator circuit to the arbitrary waveform generator (AWG) and the oscilloscope (OSC) are treated as transmission lines (see Fig.~\ref{fig2}b, and see the SI for the details of the transmission line geometry LTspice simulations). Already a $0.25\,$m long cable between the high-impedance input oscilloscope and the oscillator circuit yields a severe frequency limitation (see the deviation of the pink and black curves in Fig.~\ref{fig2}a) due to the delayed return of the signal being reflected at the oscilloscope. On the other hand, an impedance-matched $50\,\Omega$ input of the oscilloscope would result in shunting of the oscillator circuit, which would completely disable the oscillating operation. 

\subsection{Oscillator circuit optimized for high-frequency operation}
To avoid the above difficulties in the high frequency regime, further on we apply the modified, in-line oscillator scheme of Fig.~\ref{fig2}c, where the impedance-matched oscilloscope measures the current signal transmitted through the oscillator circuit ($I=V_\mathrm{OSC}/50\Omega)$. In this case no signal is reflected at the AWG and the oscilloscope, i.e. the lengths of the transmission lines between the oscillator circuit and the AWG or OSC are not relevant for the frequency of the oscillation. As a price, we cannot measure the voltage drop on the VO$_2$ element, just a signal proportional to the current is detected. On the other hand, the finite distance $d$ between the series resistor and the VO$_2$ sample becomes a crucial factor. This part of the circuit is also modeled as a transmission line.

In this optimized circuit arrangement the simulated capacitance-dependent oscillation frequencies  readily reach values above 100~MHz for $d=6.5\,$mm distance, which can be easily realized with a memristor chip and an SMD resistor on a printed circuit board (see red trace in Fig.~\ref{fig2}a and red-framed inset). Even higher, $1\,$GHz oscillation frequencies are predicted once the memristor-resistor distance is reduced to the $d=1.5\,$mm range, which arrangement, however, already prefers an integrated on-chip resistor (see blue trace in Fig.~\ref{fig2}a and blue-framed inset). 



\begin{figure}[b!]
	\centering
 \includegraphics[width=0.45\textwidth]{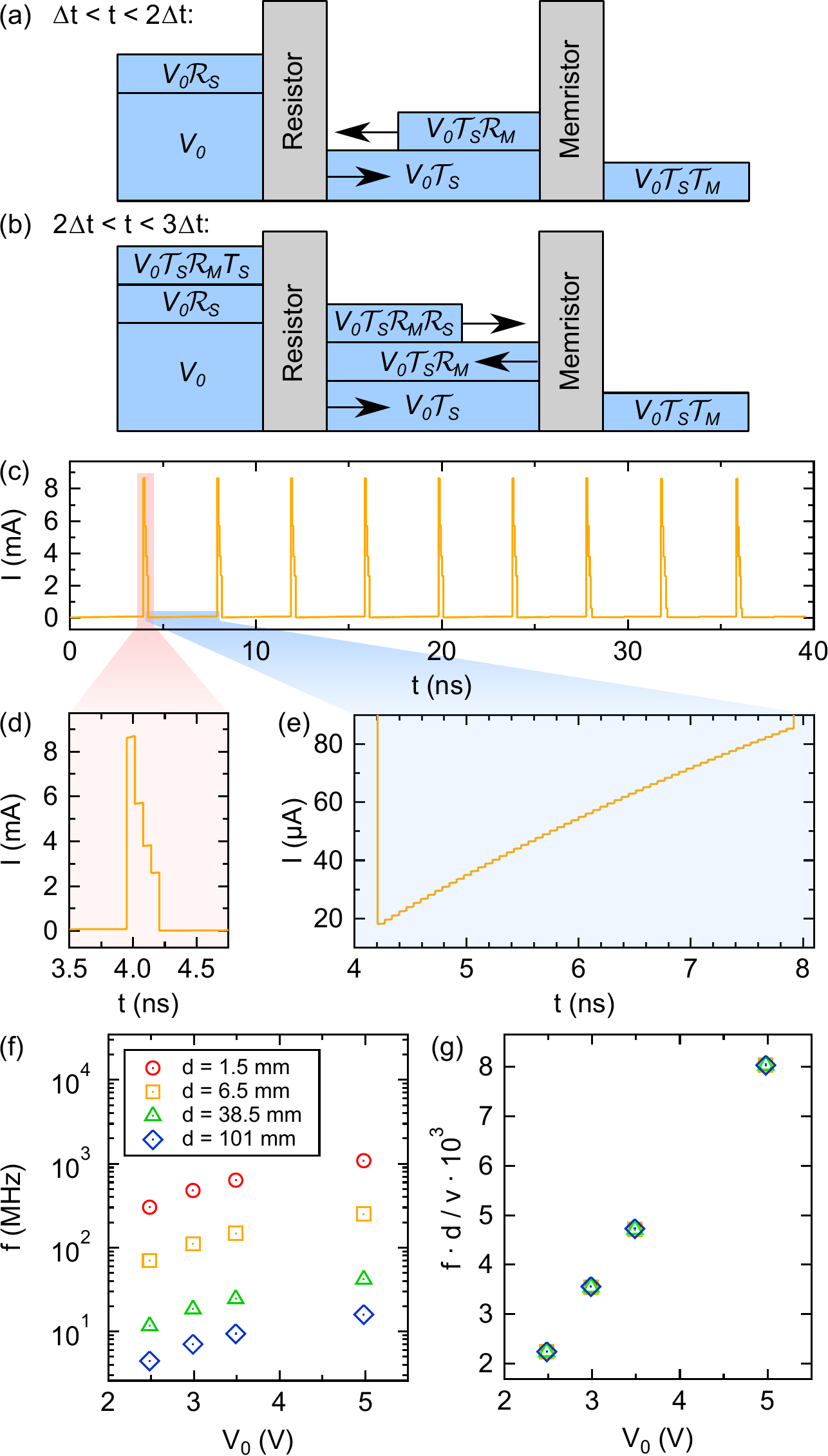}
	\caption{
    \it \textbf{Simulated oscillating time-traces in the high-frequency domain, and the scaling of the oscillation frequencies with the memristor-to-resistor distance.} (a,b) Illustration of the voltage-build-up in the circuit. At time $t=0$ a step-function excitation $0\rightarrow V_0$ arrives to the resistor in series. Panel (a) illustrates the $\Delta t <t<2\Delta t$ time interval, when the signal transmitted through the resistor has already arrived at the memristor and has been partially reflected by it, but has not yet returned to the resistor. Panel (b) similarly illustrates the $2\Delta t <t<3\Delta t$ time interval. (c) Example simulated oscillating current signal with 251~MHz oscillation frequency. (d,e) Magnified segments of panel (c) illustrating the step-wise current evolution along the reset (d) and set (e) transition. 
    (f) Simulated oscillation frequencies as a function of the $V_0$ oscillator drive voltage with different markers displaying different memristor-to-resistor distances (see legends). (g) Once the oscillation frequencies are normalized to the $1/\Delta t = v/d$ inverse signal propagation time, the frequency values demonstrated in panel (f) collapse to a single curve. The same simulation parameters are used as in Fig.\ \ref{fig2}.}
	\label{fig3}
\end{figure}

Obviously, the highest oscillation frequencies occur at the lowest capacitance values. For this reason, in the following paragraphs we analyze the operation of the predicted circuit in the absence of a parallel capacitor $C$, by considering only the frequency limitation caused by the finite memristor-to-resistor distance.
The results of these simulations are shown in Fig.~\ref{fig3}f, with the different colors showing the simulated oscillation frequencies at different distances $d$ (see legends), while the horizontal axis shows the variation of the oscillation frequencies with the constant oscillator drive voltage $V_0$. Note that here $V_0$ is the amplitude of the $0\,\mathrm{V}\rightarrow V_0$ voltage step (see the red illustration in Figure ~\ref{fig2}c) that the AWG outputs to the coaxial line with $50\,\Omega$ wave impedance, which practically results in a $2V_0$ DC voltage in the coaxial line due to the large reflection from the series resistor. These simulations show that an oscillator with a $d=6.5\,$mm of memristor-to-resistor spacing can produce frequencies in the range of our highest experimentally observed oscillations (see the $167\,$MHz oscillation in Fig.~\ref{fig1}d) when operated with a sufficiently high drive voltage $V_0$. As $d$ is increased or $V_0$ is decreased, the oscillation frequencies gradually decrease. However, if $d$ is further reduced, the predicted oscillation frequencies become even higher, but as will be shown later, this regime is subject to further experimental limitations.

Note that the above $100\,$MHz-range frequencies at $d=6.5\,$mm memristor-to-resistor distance (yellow squares in Fig.~\ref{fig3}f) correspond to a $\Delta t=d/v\approx 32\,$ps signal propagation time from the resistor to the memristor, where $v\approx2\cdot 10^8\,$m/s is the signal propagation speed in the transmission lines. This short propagation time seemingly contradicts the two orders of magnitude larger period time of the oscillation at $\approx 100\,$MHz frequency.
This apparent discrepancy is resolved by a detailed calculation of the voltage build-up on the memristive element. We assume that at time $t=0$ the voltage changes from zero to a constant $V_0$ value on the AWG and analyze how this signal propagates through the circuit (see Figs.~\ref{fig3}a,b). This step-function first arrives to the resistor in series ($R_S$), where it is partially transmitted (reflected) with a transmission coefficient $\mathcal{T}_S$ (reflection coefficient $\mathcal{R}_S$). The reflected signal is fed back to the AWG where it is absorbed by the $50\,\Omega$ output impedance. The transmission and reflection coefficients are calculated from the solution of the telegrapher's equations as 
\begin{equation}
    \mathcal{T}_S=1-\mathcal{R}_S=2Z_0/(R_S+2Z_0),
    \label{eq:transR}
\end{equation} 
where $Z_0=50\,\Omega$ is the wave impedance of the transmission lines.\cite{Csontos2023} The transmitted signal is then propagated to the memristor, where the signal is partially transmitted (reflected) with a transmission coefficient $\mathcal{T}_M$ (reflection coefficient $\mathcal{R}_M$). These coefficients are obtained from the actual $Z_M$ memristor impedance  as 
\begin{equation}
    \mathcal{T}_M=1-\mathcal{R}_M=2Z_0/(Z_M+2Z_0),
    \label{eq:trans}
\end{equation}
where $Z_M$ can be replaced by the $R_M$ memristor resistance due to the negligible capacitive impedance compared to the resistance over the frequency range of the measurement. This is underpinned by the measured $C=2\,$fF stray capacitance of our devices,\cite{doi:10.1021/acsnano.4c03840_Sebastian_PicojouleSwitching} which is significantly smaller than the $\approx 55\,$fF capacitance, at which we would start to expect signal distortion at our typical devices resistances and our $1\,$GHz measurement bandwidth (see Section S1 of the Supporting Information for the more detailed analysis of the stray capacitance).
The partial signal wave transmitted through the memristor is eventually propagated to the OSC, where it is absorbed by the $50\,\Omega$ input impedance. Due to the large series resistance $R_S$, only a very small fraction of the voltage $V_0$ reaches the memristor in the first turn, i.e. a large number of back and forth bounces are required to reach the $V_\mathrm{set}$ voltage drop  on the memristor where the set transition occurs. This voltage build-up time can be calculated analytically as follows:   
\begin{equation}
        \tau_{0\rightarrow V\mathrm{set}}=   \frac{d}{v}\left[ \frac{\ln \left(1- V_\mathrm{set} \cdot \frac{1- \mathcal{R}_M\mathcal{R}_S }{ 2V_{0} \mathcal{T}_S \mathcal{R}_M} \right)}{\ln \left( \mathcal{R}_M\mathcal{R}_S)\right)}-1
        \right]
          \label{eq:buildup}
\end{equation}  
(see Section S2 of the Supporting Information for the derivation of Eq.~\ref{eq:buildup}). Applying experimentally reasonable values of $d=6.5\,$mm, $R_S=22\,\mathrm{k}\Omega$, $R_\mathrm{M,OFF}=28\,\mathrm{k}\Omega$, $V_0=5\,$V and $V_\mathrm{set}=2.4\,$V, {a voltage build-up time of $2.2\,$ns is obtained, which  corresponds to a signal bouncing back and forth $68$ times between the resistor and the memristor.}
The above analytical formula describes the voltage build-up from zero to $V_\mathrm{set}$, but similar voltage build-up times are also delivered by the LTspice simulations of a periodic oscillation (see Fig.~\ref{fig3}c). The magnified figures of the reset (Fig.~\ref{fig3}d) and set (Fig.~\ref{fig3}e) transitions clearly demonstrate the step-wise voltage build-up, where the number of discrete steps correspond to the number of bounces between the resistor and the memristor. Fig.~\ref{fig3}d shows a rather fast reset process after the set transition. This is related to the low memristor resistance, which allows a fast release of the voltage across the memristor. However, after the reset process, the next set transition requires a large number ($\approx 57$) of back and forth bounces, as shown in Fig.~\ref{fig3}e. This explains why the oscillation frequency is eventually much lower than the $1/\Delta t \approx 31.25\,$GHz inverse propagation time. 

Finally, it is important to recognize another crucial conclusion from these simple model calculations. By increasing the distance $d$ and the corresponding propagation time $\Delta t=d/v$, we slow down the oscillation curves linearly along the time axis only. This means that a normalized, unitless $t\cdot v/d$ time axis, or the equivalent normalized $f\cdot d/v$ frequency axis, result in the collapse of the data of different distances $d$ into a single curve. This is shown in Fig.~\ref{fig3}g, where the such normalized frequency values indeed collapse on the same curve showing the same dependence on $V_0$. This property can be clearly seen from Eq.~\ref{eq:buildup} as well, where the voltage build-up time scales linearly with $\Delta t = d/v$, and the rest of the equation does not depend on $d$.

\subsection{Experimental observation of ultrafast oscillations}

\begin{figure}[b!]
	\centering
 \includegraphics[width=\linewidth]{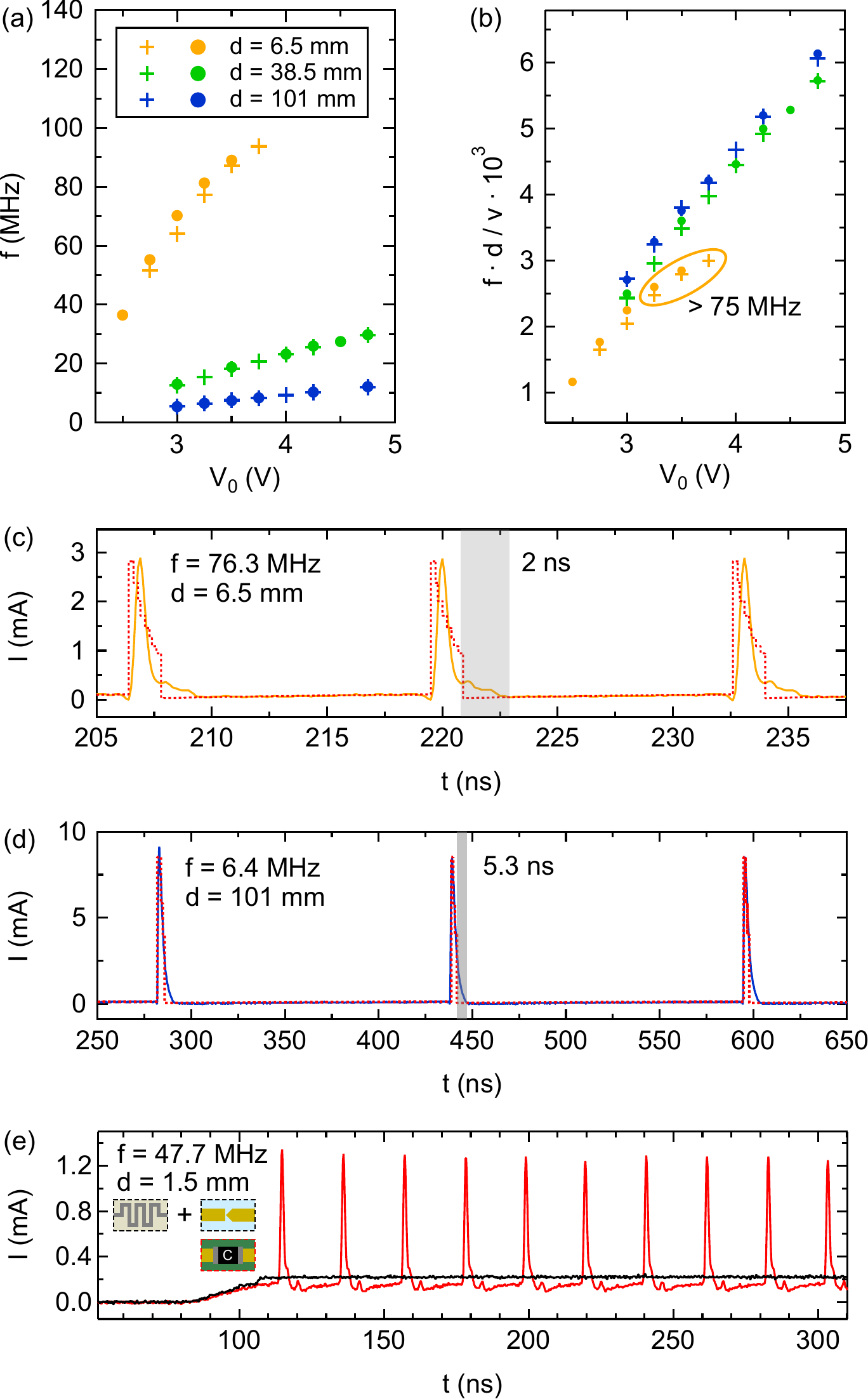}
	\caption{\it \textbf{Experimentally observed ultrafast oscillations}. (a)  Measured oscillation frequencies at various memristor-to-resistor distances ($d$, indicated by the legend) and drive voltages ($V_0$) are displayed. We plotted two measurement cycles, marked with cross and circle symbols, to demonstrate reproducibility. In one cycle, the voltage is increased at a given distance; then, the distance is increased to the next value. After measuring at all distances in the first cycle (marked with crosses), the whole procedure was repeated (marked with circles). 
    (b) Once the oscillation frequencies are normalized to the $1/\Delta t = v/d$ inverse signal propagation time, the frequency values demonstrated in panel (a) collapse to a single curve similarly to the simulations (Figs.~\ref{fig3}f,g). This scaling only fails at the highest frequencies above 75~MHz (see orange encircled measurements). (c,d) Measured oscillating time-traces (solid lines) in comparison to simulated time-traces with similar characters (red dotted lines). The experimental traces exhibit an extended tail region during the reset process, as illustrated by the gray shaded areas with the indicated widths. The distances $d$ and the oscillation frequencies $f$ are indicated in the panels. (e) Investigation of oscillator circuits with integrated, meander-shaped resistor in series. The meander's center is at $190\,\mu$m distance from the device active region, while the path length from the meander's center to the device active region is $1.5\,$mm. The black curve illustrates the measurement without a parallel capacitor: in this configuration, integrated samples with ultra-small $d$ systematically fail to oscillate (see black curve). With the insertion of a $C=1\,$pF parallel capacitor the oscillatory behavior is recaptured (red curve). 
    }
	\label{fig4}
\end{figure}

Equipped with the above finite signal propagation speed simulations, let us explore the experimental possibility of ultrafast oscillations. We have built oscillator circuits using our nanosized VO$_2$ samples and an $R_S=22\,\mathrm{k}\Omega$ SMD series resistor assembled along a transmission line on a printed circuit board (see the details of the experimental setup in the Methods section). The series resistors were placed at distances $d=6.5, 38.5, 101\,$mm from the memristor, and the respective time-traces (solid lines in Figs.\ \ref{fig4}c,d) and the related oscillation frequencies (Figs.\ \ref{fig4}a,b) were measured. These experimental data exhibit great similarities to our above discussed simulations. First, the time-traces show a similar asymmetric pattern, where the system stays shortly in the ON state, but after the reset 
process, the next set transition requires a long time compared to the width of the sharp peaks, where the system stays in the ON-state. Second, the dependence of the oscillation frequencies on $d$ and $V_0$ (Fig.~\ref{fig4}a,b) show similar trends as the simulated oscillations (Fig.~\ref{fig3}f,g). In particular, after normalizing the frequencies ($f\cdot d/v$), the measurements at different distances mostly fall on the same curve (Fig.\ \ref{fig4}b). However, at the highest frequencies this scaling fails (see the deviation of the orange datapoints from the rest of the data at higher $V_0$ values, i.e. the orange encircled measurements). Note that all these significantly deviating datapoints correspond to $>75\,$MHz oscillation frequencies. 
This implies that for the fastest oscillations the frequency is not limited by the memristor-resistor distance, rather some other time-scale, like the internal relaxation time of the memristor becomes important. The same tendency is also seen in Fig.\ \ref{fig4}c,d, where the measured time-traces (solid lines) are compared to red simulated traces with similar operation characters. This comparison highlights a clear difference at the falling edges after the current peaks: whereas the simulations always show a sharp reset, in the experiment the falling edges exhibit extended tails, which we also attribute to an internal relaxation time of the memristor. These tails extend over the time-scale of a few nanoseconds, as illustrated by the light gray shaded areas. This phenomenon becomes important if the oscillation period is comparable to the duration of this tail region.  

The previous observations indicate the interplay of geometry-induced voltage build-up times and internal physical relaxation times in determining the fastest possible oscillation frequencies. From this point of view it is interesting to explore devices, where the memristor-to-resistor distance is even smaller. To this end, we have fabricated memristor devices with integrated meander-shaped resistor in series yielding $d\approx 1.5\,$mm (see the Methods section for more details).  For  such short distance the simulations in Fig.~\ref{fig2}a predict $>1\,$GHz oscillation frequencies. As a sharp contrast, for such spatially confined devices we have systematically not observed oscillatory behavior, rather the system reaches a steady state when the $V_0$ voltage is applied (black line in Fig.~\ref{fig4}e). If, however, the device operation is artificially slowed down by a $C=1\, \rm pF$ parallel capacitor, the oscillating operation can be recovered (red line in Fig. \ref{fig4}e). 
We explain this as follows. Either a properly chosen parallel capacitor, or a properly distant resistor in series helps to slow-down the voltage variation in the circuit, such that after reaching the set (reset voltage) these voltage levels are kept for a while to ensure a complete set (reset) transition. Without this artificial slow-down of the voltage variation, the system would not have enough time to complete the set (reset) transition, and the system would stick to some stable state instead of an oscillation.
To support the above explanation, we constructed Matlab Simulink simulations where the set and reset switching timescales of the memristors are modeled through equations for the time derivative of the device resistance. These simulations reproduce the experimentally reached 167~MHz frequency maximum (see Section S3 of the Supporting Information), and also illustrate the role of delayed voltage build-up in stabilizing the oscillation. The latter is demonstrated in Sections S4 and S5 of the Supporting Information conclusively reproducing the phenomenon seen in Fig.~\ref{fig4}e.
According to the above arguments, the fastest oscillation is achieved when the internal relaxation time of the memristor matches the time-scale at which the finite $d$ distance or the $C$ parallel capacitor keeps the voltage close to the set or reset voltage value once these are reached along the oscillation. 

Following these considerations, we investigate the realistic internal relaxation time-scales of our devices, which play a fundamental role in determining the maximum possible oscillation frequency.

\subsection{Investigation of the internal relaxation time-scales}

Using VO$_2$ memristors with the same geometry as the one tested here, we have demonstrated set times below $15\,$ps and reset times below $600\,$ps \cite{doi:10.1021/acsnano.4c03840_Sebastian_PicojouleSwitching}, which would allow oscillation frequencies up to above GHz. However, these shortest switching time values were achieved under optimized conditions. The $15\,$ps set time was demonstrated with an ultrashort, $20\,$ps FWHM switching pulse of amplitude significantly exceeding the set threshold voltage, while the $600\,$ps reset time was achieved with the least invasive measurement, i.e., the set transition was performed with the lowest possible meaningful set pulse, after which the voltage was immediately taken off the sample and the time it takes for the system to relax to the OFF-state was scanned with ultra-short ($20\,$ps) readout pulses. 

In what follows, we argue that significantly longer set/reset times than these can be experienced under less optimized conditions, which are unavoidable during oscillating operation. Most importantly, in VO$_2$ oscillator circuits, the maximum and minimum voltages available are mainly determined by the set voltage and the reset voltage, as explained below. As soon as the set voltage is reached on the VO$_2$ device, the resistance starts to decrease (set transition), so the device voltage also starts to decrease due to the voltage division with the series resistor. Likewise, as the voltage decreases to the reset voltage, the resistance of the sample starts to increase (reset transition), so the voltage across the sample also starts to increase. Although minor overshoots or undershoots are possible due to the finite voltage build-up time, in principle, no voltages significantly higher (lower) than $V_\mathrm{set}$ (and $V_\mathrm{reset}$) can be obtained on the VO$_2$ sample during oscillation.

In an oscillator circuit, however, the driving conditions are constantly changing, making it difficult to directly study the switching times and their dependence on the driving parameters.  To overcome this difficulty, we prefer to investigate the switching times by pulsed measurements, where the driving conditions are well controlled, while the pulsed current variation still somewhat resembles the oscillating time-traces. Specifically, oscillatory operation is mimicked by using a series of pulses whose amplitude is close to the set transition, while the constant readout voltage between the pulses represents a level close to the reset transition.


\begin{figure}[b!]
	\centering
 \includegraphics[width=0.48\textwidth]{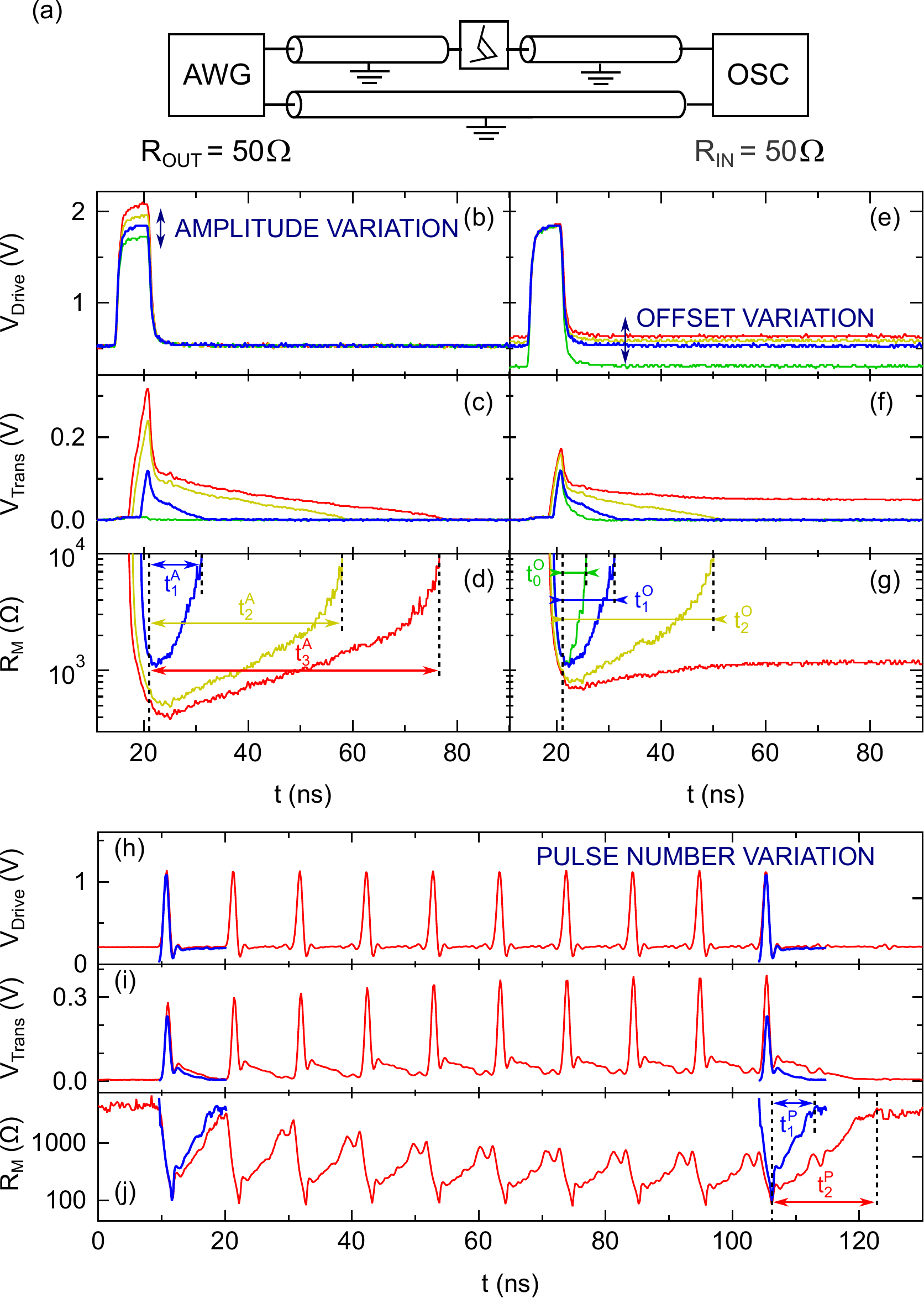}
	\caption{\it \textbf{Investigation of the physical relaxation times by pulsed experiments.} (a) The scheme of the measurement setup. (b) Driving pulses for our experiments with variable pulse amplitude and fixed readout voltage offset, with pulse amplitudes of 1.7~V (green curve), 1.85~V (blue curve), 1.95~V (yellow curve), 2.1~V (red curve). (c) Transmitted voltage through the VO$_2$ sample as measured by the oscilloscope. The time axis of panels (b) and (c) are aligned to eliminate the finite propagation times in the coaxial lines connecting to the AWG and OSC. (d) The memristor resitance $R_M(t)$ evaluated according to Eq.~\ref{eq:RMt}. (e,f,g) Driving pulses for our experiments with fixed pulse amplitude and variable readout voltage offsets of 0.31~V (green curve), 0.53~V (blue curve), 0.59~V (yellow curve), 0.64~V (red curve), (e) together with the corresponding transmitted voltage pulses (f), and the calculated $R_M(t)$ traces (g). (h,i,j) Experiments with single driving pulses (blue) and ten times repeated driving pulses (red) (h) together with the corresponding transmitted voltage pulses (i), and the calculated $R_M(t)$ traces (j). For comparison, the single pulse measurements (blue) are also replotted at the time of the tenth repetitive pulse (red). 
    }
	\label{fig5}
\end{figure}

The scheme of our measurement is sketched in Fig.~\ref{fig5}a. In these experiments the AWG outputs ultrashort voltage pulses ($V_\mathrm{Drive}(t)$) instead of a step function, which are measured directly by one channel of the oscilloscope. The same pulses are driving the memristor sample from the other output channel of the AWG, and the pulses transmitted through the sample ($V_\mathrm{Trans}(t)$) are measured at another channel of the OSC (see further details of the experimental setup in the Methods section). In this scheme, no resistor in series is applied, but the switching dynamics of the memristor sample alone is tested. 

Note that the transmission coefficient in Eq.~\ref{eq:trans} would yield a frequency-dependent transmission for a capacitive or inductive $Z_M$ memristor impedance, but for a dominantly resistive $Z_M=R_M$, which is the case in our measurements (see discussion after Eq.~\ref{eq:trans}), the transmission coefficient is frequency independent. This property ensures that an incoming wave-package, such as a pulse, preserves its shape along the transmission, only its amplitude is modified, as long as $R_M$ is constant in time. Or, inversely, the temporal variation of the memristor resistance can be directly traced from the ratio of the transmitted and incoming (driving) signals:\cite{doi:10.1021/acsnano.4c03840_Sebastian_PicojouleSwitching}
\begin{equation}
   \label{eq:RMt}
    R_M(t)= 2Z_0\cdot\left(\frac{V_\mathrm{Drive}(t)}{V_\mathrm{Trans}(t)}-1\right).
\end{equation}
Note, that the transmitted voltage signal is directly proportional to the memristor current as $V_\mathrm{Trans}(t)=I(t)\cdot Z_0$.
Figs.~\ref{fig5}b,c,d,  Figs.~\ref{fig5}e,f,g and Figs.~\ref{fig5}h,i,j respectively demonstrate the $V_\mathrm{Drive}(t)$ driving waveforms output from the AWG (top panels), the $V_\mathrm{Trans}(t)$ transmitted waveforms (middle panels) measured at the OSC and the $R_M(t)$ temporal variation of the memristor resistance calculated from Eq.~\ref{eq:RMt} (bottom panels) for various driving signals. The bottom panels are cut at $10^4\,\Omega$, a resistance slightly below our resolution limit due to the finite 8 bit resolution of our OSC. 

In Figs.~\ref{fig5}b,c,d and Figs.~\ref{fig5}e,f,g we investigate the response of the memristive sample to single driving pulses with variable amplitudes and readout offset voltages. 
Our primary focus is on the blue curve (the same in the right and left panels), where the pulse amplitude and offset are adjusted close to the set and reset voltage. The green curves in Figs.~\ref{fig5}b,c show that a slightly smaller pulse amplitude no longer turns the sample on, while the red curves in Figs.~\ref{fig5}e,f,g demonstrate that by applying a slightly higher readout voltage, the system is already stuck in the ON state without turning OFF.

To define a comparable measure for the duration of the set and reset transitions at the different driving conditions, we measure the set time as the time between the middle of the rising edge of the drive pulse and reaching $<2\,\mathrm{k}\Omega$ resistance, while the reset time is the time between the middle of the falling edge of the drive pulse and reaching $>10\,\mathrm{k}\Omega$ resistance. The latter relaxation times are denoted by the correspondingly colored arrows in the figure. For the blue reference measurement the above definitions yield a set time of $5.4\,$ns and a reset time of $t_1^A=t_1^O=9.6\,$ns.

Next, we investigate how these times vary by changing the pulse amplitude and the readout offset voltage. Once the pulse amplitude is increased compared to the blue curve (see the yellow and red curves in Figs.~\ref{fig5}b,c,d) the set time respectively decreases to  $3.7\,$ns and  $2.8\,$ns, while the reset time significantly increases to $t_2^A=36.4\,$ns and $t_3^A=54.4\,$ns, respectively. On the other hand, the minor increase of the readout offset voltage (yellow and red curves in Figs.~\ref{fig5}e,f,g) yields minor variation of the set time, while the reset time shows a significant increase ($t_2^O=28.8\,$ns for the yellow curve, and $\gg 100\,$ns for the red curve). We also tested a significantly lower readout voltage than the blue curve (green curve), which reduces the reset time to $t_0^O=4.4\,$ns. The latter value is still much longer than the fastest $600\,$ps relaxation at zero readout voltage,\cite{doi:10.1021/acsnano.4c03840_Sebastian_PicojouleSwitching} while the readout voltage corresponding to the green curve is already unrealistically small for an oscillator operation. This means that the above $t_0^O=4.4\,$ns reset time is a lower estimate of the relaxation time achievable under oscillator conditions in this particular case.

Furthermore, the relaxation time is also very sensitive to the number of applied driving pulses (Figs.~\ref{fig5}h,i,j). In these experiments, a relaxation time of $t_1^P=6.8\,$ns is measured after a single $\approx 1.1\,$ns long driving pulse (blue curves), while the ten-times repetition of the same driving pulse with $10.6\,$ns period time (red curves) yields a relaxation time of $t_2^P=16.6\,$ns after the last pulse. Here, the single blue pulse is replotted at the time of the 10$^\mathrm{th}$ pulse for comparison.

\begin{figure}[b!]
	\centering
    \includegraphics[width=0.42\textwidth]{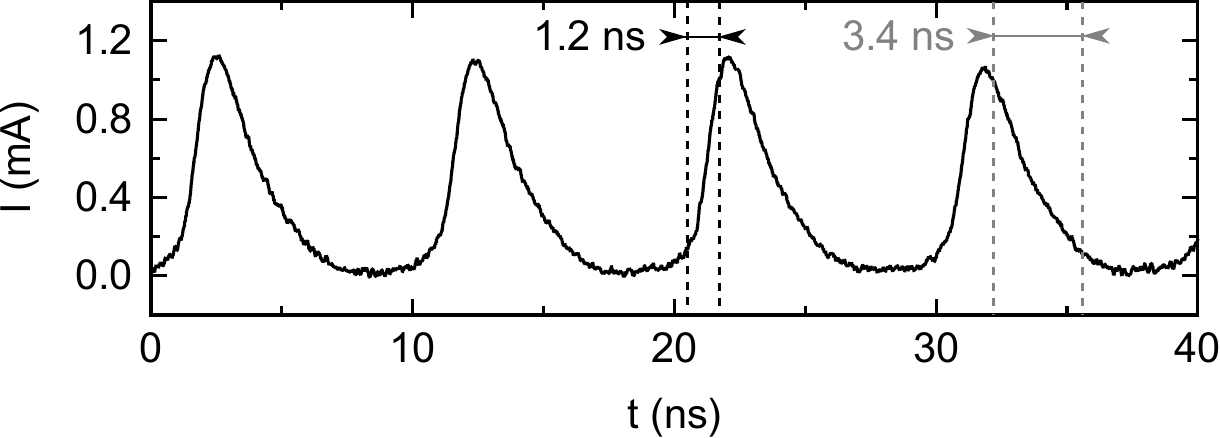}
	\caption{\it \textbf{Ultrafast oscillations measured by our $100\,$GHz bandwidth setup.} The measurements are performed by the experimental setup used in Ref.~\citenum{doi:10.1021/acsnano.4c03840_Sebastian_PicojouleSwitching} and an $R_S=16\,\mathrm{k}\Omega$ resistor
    .
    }
	\label{fig6}
\end{figure}

Finally, we gain a better insight into the time scale of the set transitions during oscillator operation by testing the oscillator circuits with our $100\,$GHz bandwidth measurement setup, which we also used for the ultra-short switching time experiments.\cite{doi:10.1021/acsnano.4c03840_Sebastian_PicojouleSwitching} The such-measured oscillating signal in Fig.~\ref{fig6} yields similar $103\,$MHz oscillation frequency, respectively exhibiting a $10\%$ to $90\%$ rise-time of $1.2\,$ns and fall-time of $3.4\,$ns for the set and reset transition. This confirms that the reset transition is the dominating limiting factor. On the other hand, the observed $1.2\,$ns rise time is significantly longer than the $15\,$ps set time measured in our previous work.  \cite{doi:10.1021/acsnano.4c03840_Sebastian_PicojouleSwitching} This $100\,$GHz bandwidth measurement clearly demonstrates that the $1.2\,$ns rise time is not an instrumental bandwidth limitation, but the set time is really elongated. This is attributed to the fact that in an oscillator the voltage is limited by the set voltage, while the $15\,$ps set time was observed at pulsed driving with significantly higher pulse amplitudes. We also note, that the $1.2\,$ns rise time is shorter than the several nanosecond long set times observed in Figs.~\ref{fig5}b,c,d for pulses adjusted close to the set transition. This apparent difference is attributed to the fact that for single pulses the switching starts from a room temperature, fully developed OFF state. However, during oscillator operation the active region heats up slightly, the reset transition is incomplete, and thus a faster set transition is achieved than for single pulse driving.

The above illustrative experiments on the switching dynamics of VO$_2$, as well as previous sub-threshold firing experiments and further VO$_2$ dynamics studies~\cite{delValle2019,Schuller2021} show that VO$_2$ devices cannot be treated as simple hysteresis switches with well-defined switching voltages and switching times. On the contrary, the switching parameters depend sensitively on the drive parameters and the history of the device, with switching time variations of up to an order of magnitude. A comprehensive analysis of the full relaxation dynamics and the device-to-device variation of the transition times is clearly beyond the scope of this paper.  Nevertheless, our above switching-time experiments and our similar experiments on other devices already point to the following conclusions:
(i) Under the conditions of an oscillator circuit, i.e., for repetitive voltage oscillations not significantly exceeding the set voltage and not going below the reset voltage, the set and reset times are expected to be order(s) of magnitude larger than the minimum set time of $15\,$ps and relaxation time of $600\,$ps. (ii) The reset transition is always significantly longer than the set transition, i.e. the relaxation time is the key restricting factor for the oscillation frequency. (iii) The observed relaxation times make it clear that an oscillation significantly faster than the fastest $167\,$MHz oscillation achieved is not realistic. Our experience with these ultrafast oscillator circuits shows that oscillations in the $75-100\,$MHz regime are routinely established. However, even larger frequencies, like the $167\,$MHz oscillation in Fig.~\ref{fig1}d, specifically rely on the fine interplay of the device parameters, which is not satisfied in every oscillator circuit.

\section{Conclusions}
In conclusion, we have demonstrated VO$_2$ oscillator circuits optimized for high-frequency, $100\,$MHz-range operation. To this end, we have applied an optimized sample layout, where the stray capacitance is minimized, and the switching is focused to an ultrasmall, $\approx 30\,$nm wide active region. Furthermore, the circuit layout is also optimized for high-frequency operation by applying a transmission line geometry, where the oscillation frequency is tunable both by the $d$ memristor-to-resistor distance and the $C$ parallel capacitance. According to circuit simulations of the same transmission-line geometry, reasonably achievable, pF-range parallel capacitances and a few mm memristor-to-resistor distances readily yield $100\,$MHz-range oscillation frequencies, similarly to our experimentally demonstrated oscillations up to $167\,$MHz frequency. 

Circuit simulations would allow even higher $>\,$GHz oscillation frequencies with even shorter $d$ and lower $C$. At the same time the fastest pulsed switching experiments demonstrate subnanosecond relaxation times,\cite{doi:10.1021/acsnano.4c03840_Sebastian_PicojouleSwitching}
which could also be compatible with GHz oscillator operation. However, we have demonstrated that the speed limitation is fundamentally different for an oscillator circuit than for optimized resistive switching experiments with single pulses.
Most importantly, the finite $\gtrsim V_\mathrm{reset}$ voltage level on the memristive element and the rapidly repeated switching events lead to a slowdown of the relaxation time, which explains the $100\,$MHz-range frequency limit achieved. Together with these constraints, the above results have enabled oscillation frequencies more than an order of magnitude higher than the fastest VO$_2$ oscillators presented so far,\cite{10.1063/1.4922122_9MHz_cikk} paving the way towards the realization of ultra-fast and energy-efficient VO$_2$-based oscillating neural networks.

\section{Methods}

\subsection{Sample fabrication}

The vertical layer structure of our VO$_2$ devices is shown in {Fig.~\ref{fig1}e} including the Al$_2$O$_3$ substrate, the $50\,$nm thick epitaxial VO$_2$ layer, a $5\,$nm Ti adhesive layer and the $60\,$nm thick gold top electrodes. The VO$_2$ layer was deposited on the substrate by pulsed laser deposition method according to Ref.~\citenum{Kim2017}. The top electrodes were patterned by standard electron beam lithography and deposited by electron-beam evaporation at 10$^{-7}$~mbar base pressure at rates of 0.1~nm/s (Ti) and 0.4~nm/s (Au), followed by lift-off.

Small, $\approx1.5$~mm memristor-to-resistor distances were achieved by samples, where a meander-shaped series resistor was integrated on the memristor chip (see the illustrative inset in Fig.~\ref{fig2}a). The line width of the meanders was $2~\mu$m with $2~\mu$m spacing, $150~\mu$m length (for a single line) and $N=5$, 9, 13, 18 turns. The meanders were located at $150~\mu$m distance from the memristors. For the calculation of an effective memristor-resistor distance, the total length of the meander lines has to be considered, which can be estimated as $150~\mu\text{m}\cdot18/2 + 150~\mu\text{m} = 1.5~$mm for the longest meander with $N=18$ turns.

The VO$_2$ layer beneath the meanders was etched away in a reactive ion etching process using a Diener low-pressure plasma system. The etching mask was created in a standard electron beam lithography process. First, the vacuum chamber was pumped to reach 0.16~mbar base pressure, then pure CH$_4$ gas was supplied with 42 sccm flow rate to set the 0.64~mbar process pressure. The plasma power was set to 240~W and the duration of the etching to 30~s. Afterwards, the meander resistors were patterned by electron beam lithography, and 15 nm Pt was deposited by electron beam evaporation using 0.3 nm/s rate, followed by lift-off.

\subsection{Measurement scheme}
The oscillating traces of the VO$_2$ circuits were investigated with a Rohde \& Schwarz RTO1024 oscilloscope with $1\,$GHz bandwidth and $10\,$GS/s sampling. The drive voltage was supplied by an Agilent 33210A arbitrary waveform generator. In order to avoid sample degradation due to too many oscillation periods, the constant drive voltage was applied to the oscillator circuit for a finite time, typically achieving $\approx 50$ periods of oscillations per measurement. 

Most of the measurements were performed on $3.0 \,\mathrm{cm}\times 2.4\,\mathrm{cm}$ printed circuit boards with SMA connectors on both sides, and $50\,\Omega$ wave impedance transmission lines connecting the SMA connector to the surface mounted series resistor, the resistor to the bonding wires of the sample, and the sample to the other SMA connector. 

{For the pulsed measurements in Fig.~\ref{fig5} either a Zurich Instruments HDAWG or an AT1120 AWG module mounted on a National Instruments PXIe-7976 board together with a Mini-Circuit ZHL-72A+ amplifier were applied. }

The $\approx 100\,$GHz bandwidth measurements in Fig.~\ref{fig6} were performed with the same setup as the measurements in Ref.~\citenum{doi:10.1021/acsnano.4c03840_Sebastian_PicojouleSwitching}. The samples were contacted by two 67 GHz bandwidth Picoprobe triple probes in a probe station. A Micram DAC10004 100GSa/s DAC unit together with a Centellax UA0L65VM broadband amplifier served as the driving unit. The transmitted voltage was recorded by a Keysight UXR1104A digital storage oscilloscope at $256\,$GSa/s sampling rate and $113\,$GHz analog bandwidth. The input terminals of the oscilloscope were protected by RF attenuators. 

\subsection{Simulation}

\begin{figure}[b!]
	\centering
    \includegraphics[width=0.48\textwidth]{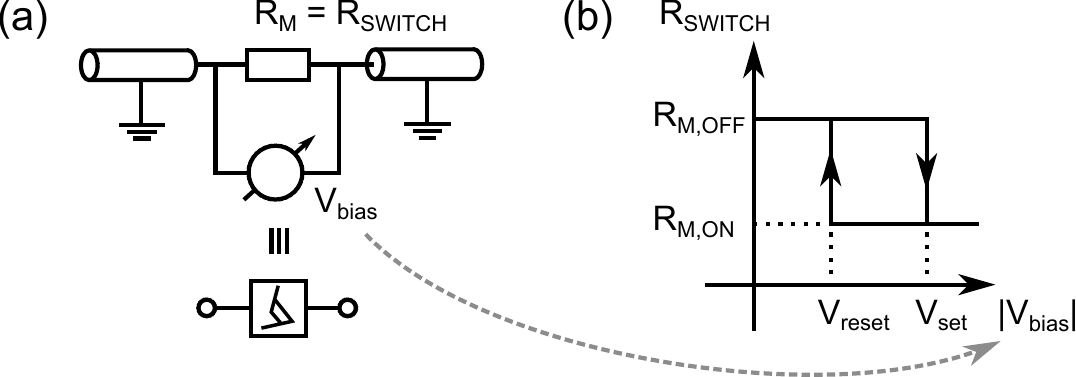}
	\caption{\it \textbf{LTspice simulations of the VO$_2$ oscillator circuits.} In most of our investigation the VO$_2$ memristor is placed in a waveguide geometry circuit, like Fig.~\ref{fig2}c. These arrangements are simulated by (i) measuring the $V_\mathrm{bias}$ voltage drop on the memristor, which acts as a variable resistor, $R_M$ (panel (a)); and (ii) updating the memristor resistance according to the variation of $V_\mathrm{bias}$ and the hysteretic resistance switch model in panel (b). The OFF-state (ON-state) memristor executes a set (reset) transition once the voltage goes above (goes below) $V_\mathrm{set}$ ($V_\mathrm{reset}$).
    }
	\label{fig7}
\end{figure}

The oscillating circuits were simulated by LTspice and Matlab Simulink. In order to place our memristor in a circuit with a waveguide geometry (Fig.~\ref{fig7}a), it was important that it acted as a variable resistor, instead of low-frequency simulations of VO$_2$ devices,\cite{Maffezzoni2015} where the memristor model contained a current generator element in the circuit. In Figs.~\ref{fig2},\ref{fig3}, as well as in the comparative analysis of Fig.~\ref{fig4}c,d (dotted lines) we have used a minimal model, where the memristor is simply a hysteretic resistance switch in LTspice, and no internal relaxation time is considered (see Fig.~\ref{fig7}b.) These simulations can track the key consequences of the finite memristor-to-resistor distance or the finite parallel capacitance. More refined Matlab Simulink simulations are presented in Sections S3-S5 of the Supporting Information, where finite and different physical relaxation times are used for the set and reset transitions. 

\section*{Author Contributions}

The LTspice simulation and experiment involving VO$_{2}$-based memristors and oscillator circuits were developed and performed by Z.P. under the daily guidance of T.N.T.. The Matlab Simulink simulations were developed by T.N.T.. The experiment carried out with the 100 GHz bandwidth setup was performed by S.W.S and M.C. in the group of J.L. Technical support for the measurements was provided by Z.B. (low frequency characterization, printed circuit board development and the high frequency probe station development) and by A.B. (fast pulsed measurements). The VO$_{2}$ memristors were developed and fabricated by T.N.T. and L.P., and the integrated meander resistor-memristor samples were designed by T.N.T. in the group of J.V.. The VO$_{2}$ thin layers were manufactured and optimized by H.K. and A.P. The project was conceived and supervised by A.H. The manuscript was written by Z.P., T.N.T. and A.H..  All authors contributed to the discussion of the results and the preparation of the manuscript.

\section*{Conflicts of interest}
There are no conflicts to declare.

\section*{Acknowledgements}

This research was supported by the Ministry of Culture and Innovation and the National Research, Development and Innovation Office within the Quantum Information National Laboratory of Hungary (Grant No. 2022-2.1.1-NL-2022-00004), and the NKFI K143169, K143282 and TKP2021-NVA-03 grants. L.P. acknowledges the support of the Bolyai J\'{a}nos Research Scholarship of the Hungarian Academy of Sciences and the University Research Scholarship Programme. J.L. and M.C. acknowledge the financial support of the Werner Siemens Stiftung. Z.P. acknowledges the support of the  Ministry of Culture and Innovation and the National Research, Development and Innovation Office within the University Research Scholarship Programme.

\bibliography{References.bib}

\end{document}


\title{VO$_2$ oscillator circuits optimized for ultrafast, 100 MHz-range operation\\ \vspace{0.4cm}
Supporting Information \vspace{0.2cm}}

\author{Zsigmond Pollner}\altaffiliation{Contributed equally to this work}
\affiliation{Department of Physics, Institute of Physics, Budapest University of Technology and Economics, M\H{u}egyetem rkp. 3., H-1111 Budapest, Hungary.\looseness=-1}
\affiliation{HUN-REN–BME Condensed Matter Research Group,
Budapest University of Technology and Economics,
Műegyetem rkp. 3., H-1111 Budapest, Hungary.\looseness=-1}

\author{Tímea Nóra Török}\altaffiliation{Contributed equally to this work}
\affiliation{Department of Physics, Institute of Physics, Budapest University of Technology and Economics, M\H{u}egyetem rkp. 3., H-1111 Budapest, Hungary.\looseness=-1}
\affiliation{Institute of Technical Physics and Materials Science,\unpenalty~Centre for Energy Research, Konkoly-Thege M. \'{u}t 29-33, 1121 Budapest, Hungary.\looseness=-1}

\author{László Pósa}
\affiliation{Department of Physics, Institute of Physics, Budapest University of Technology and Economics, M\H{u}egyetem rkp. 3., H-1111 Budapest, Hungary.\looseness=-1}
\affiliation{Institute of Technical Physics and Materials Science,\unpenalty~Centre for Energy Research, Konkoly-Thege M. \'{u}t 29-33, 1121 Budapest, Hungary.\looseness=-1}

\author{{Miklós Csontos}}
\affiliation{Institute of Electromagnetic Fields, ETH Zurich, Gloriastrasse 35, 8092 Zurich, Switzerland.\looseness=-1}

\author{{Sebastian Werner Schmid}}
\affiliation{Department of Physics, Institute of Physics, Budapest University of Technology and Economics, M\H{u}egyetem rkp. 3., H-1111 Budapest, Hungary.\looseness=-1}
\affiliation{Experimental Physics V, Center for Electronic Correlations and Magnetism, University of Augsburg, 86135 Augsburg, Germany.\looseness=-1}

\author{Zoltán Balogh}
\affiliation{Department of Physics, Institute of Physics, Budapest University of Technology and Economics, M\H{u}egyetem rkp. 3., H-1111 Budapest, Hungary.\looseness=-1}
\affiliation{HUN-REN–BME Condensed Matter Research Group,
Budapest University of Technology and Economics,
Műegyetem rkp. 3., H-1111 Budapest, Hungary.\looseness=-1}

\author{András Bükkfejes}
\affiliation{Emerson - Test and Measurement (NI), Hungária körút 30/A, H-1087 Budapest, Hungary.\looseness=-1}

\author{Heungsoo Kim}
\affiliation{Naval Research Laboratory, 4555 Overlook Ave, Washington, DC 20375, USA.}

\author{Alberto Piqué}
\affiliation{Naval Research Laboratory, 4555 Overlook Ave, Washington, DC 20375, USA.}

\author{{Jeurg Leuthold}}
\affiliation{Institute of Electromagnetic Fields, ETH Zurich, Gloriastrasse 35, 8092 Zurich, Switzerland.\looseness=-1}

\author{János Volk}
\affiliation{Institute of Technical Physics and Materials Science,\unpenalty~Centre for Energy Research, Konkoly-Thege M. \'{u}t 29-33, 1121 Budapest, Hungary.\looseness=-1}

\author{András Halbritter}
\email{halbritter.andras@ttk.bme.hu}
\affiliation{Department of Physics, Institute of Physics, Budapest University of Technology and Economics, M\H{u}egyetem rkp. 3., H-1111 Budapest, Hungary.\looseness=-1}
\affiliation{HUN-REN–BME Condensed Matter Research Group,
Budapest University of Technology and Economics,
Műegyetem rkp. 3., H-1111 Budapest, Hungary.\looseness=-1}

\maketitle
\thispagestyle{fancy}
\lhead[]{Supporting Information}
\rhead[]{}

\newpage

\setcounter{secnumdepth}{1}



\section{Investigation of the stray capacitance}

\begin{figure}[b!]
    \centering
    \includegraphics[width=0.85\linewidth]{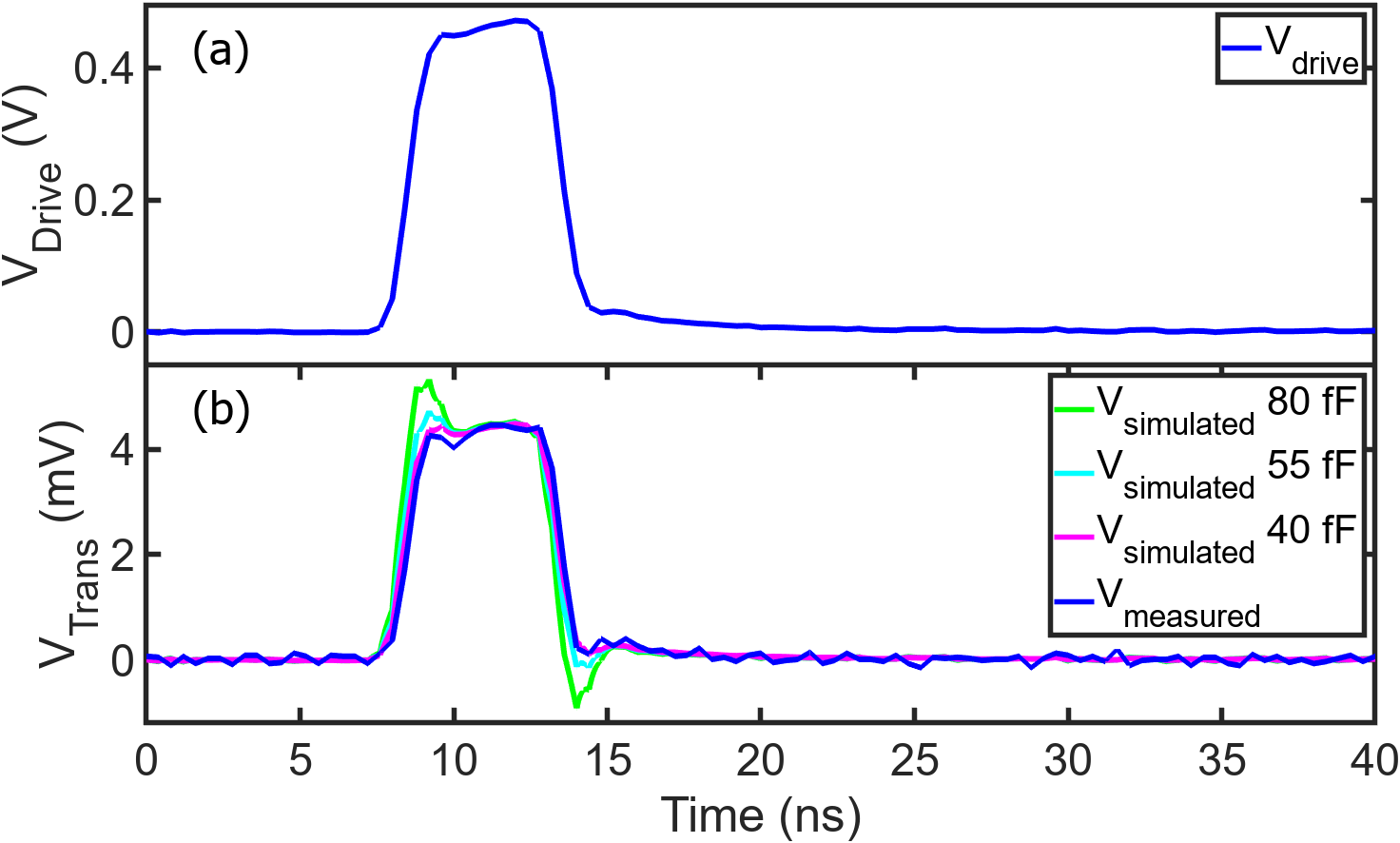}
    \caption{\it \textbf{Characterization of stray capacitance using short pulse excitation.}
    (a) Voltage pulse applied to the memristor to determine stray capacitance, with amplitude below the set threshold. (b) Measured transmitted pulse (blue) compared to simulated responses with different stray capacitance values indicated by the legend (pink, light blue and green), indicating $\lesssim 55\,$fF stray capacitance.}
    \label{fig:CapacitanceMeasurementTransientAnalysis}
\end{figure}

To estimate the stray capacitance of our setup we have excited our VO$_2$ samples by $5\,$ns long pulses applying the fastest possible rise and fall times in our $1\,$GHz bandwidth setup. The pulse amplitude was adjusted well below the set threshold voltage, so that the pulse probes the OFF state as a non-variable resistor, i.e. without any sign of switching. The blue curves in Figs.~\ref{fig:CapacitanceMeasurementTransientAnalysis}a,b respectively demonstrate the driving pulse and the pulse transmitted through the VO$_2$ sample. The pink, light blue and green lines in Fig.~\ref{fig:CapacitanceMeasurementTransientAnalysis}b demonstrate LTspice simulations for the expected transmitted signals using the same resistance as the investigated OFF resistance of the sample, but various parallel capacitances. It is clear, that parallel capacitances of $\gtrsim55\,$fF
would already cause measurable overshoots and undershoots at the pulse beginning and pulse end. The absence of these in the measured signal indicate, that the stray capacitance is smaller, and accordingly the memristor sample really acts as a resistive circuit element without any disturbing capacitive effects. It is noted, that our previous higher frequency pulsed measurements were even able to resolve ultra-small, $\approx 2\,$fF stray capacitance in our similarly prepared VO$_2$ devices~\cite{doi:10.1021/acsnano.4c03840_Sebastian_PicojouleSwitching}.

\newpage

\section{Calculation of the voltage build-up time}

The signal propagation in transmission lines is  represented by the characteristic wave impedance $Z_0$, which describes the relation between the $V^{\pm}(x,t) = V^{\pm}_0 e^{i(\omega t \pm kx)}$ voltage waves and $I^{\pm}(x,t) = I^{\pm}_0 e^{i(\omega t \pm kx)}$ current waves according to the formula $V^{\pm}_0=\pm I^{\pm}_0\cdot Z_0$. Assume, that a $R_M$ resistor is placed between two transmission lines,  and a $V_\mathrm{Drive}(t)=V_\mathrm{Drive,0}\cdot  e^{i(\omega t - kx)}$ harmonic driving signal approaches this circuit element. This driving signal is partially reflected ($V_\mathrm{Refl}(t)=V_\mathrm{Refl,0}\cdot  e^{i(\omega t + kx)}$), and partially transmitted ($V_\mathrm{Trans}(t)=V_\mathrm{Trans,0}\cdot  e^{i(\omega t - kx)}$). The current continuity relation at the $x=0$ position of the $R_M$ resistance yields the $V_\mathrm{Drive,0}-V_\mathrm{Refl,0}=V_\mathrm{Trans,0}$ condition, while Ohm's law for the $R_M$ resistor yields $V_\mathrm{Drive,0}+V_\mathrm{Refl,0}-V_\mathrm{Trans,0}=R_M\cdot V_\mathrm{Trans,0}/Z_0$. 
Combining these equations we obtain the $\mathcal{T}_M$ transmission and $\mathcal{R}_M$ reflection coefficients:
\begin{equation} \label{transzmisszio_gyorsesetben}
    \mathcal{T}_M=\frac{V_{Trans,0}}{V_{Drive,0}} = \frac{2 Z_0}{R_M + 2 Z_0},
\end{equation}
\begin{equation} \label{reflexio_gyorsesetben}
    \mathcal{R}_M= \frac{V_{Refl,0}}{V_{Drive,0}} = \frac{R_M}{R_M + 2 Z_0}.
\end{equation}
It is important to note that the above formulas are derived for a given $\omega$ angular frequency, but due to the frequency-independence of the $R_M$ resistance, the same  transmission and reflection ratios apply for all frequency components. Accordingly, the above equations are valid not only for a harmonic signal but also for other waveforms, like a pulse or a step function. This means, that e.g. a step-function wave representing a transition from zero voltage to $V_0$ is transmitted through the resistor by maintaining the shape of the wave, but reducing its amplitude from $V_0$ to $V_0\cdot 2Z_0/(R_M+Z_0)$. 

According to these introductory remarks, the voltage build-up on the series resistor memristor pair (see Fig.~3a,b in the main text) due to a step function driving is calculated as follows. The zero time ($t=0$) is defined by the moment, when the step function arrives to the series resistor, i.e. when the voltage on its left side increases from zero to $V_0$. The time of the signal propagation between the resistor and the memristor is denoted by $\Delta t$. At time $t=\Delta t$ the signal arrives to the memristor with an incoming voltage of $V_0\mathcal{T}_S$, and a reflected voltage component of $V_0\mathcal{T}_S\mathcal{R}_M$ (see the notation and description of Eqs. 1,2 in the main text).  Meanwhile, a transmitted voltage component of $V_0\mathcal{T}_S\mathcal{T}_M$
appears at the other side of the memristor. This voltage is constant until the voltage component reflected on the memristor bounces back and forth between the resistor and the memristor, but afterwards, i.e. at time $t=3\Delta t$ the incoming voltage at the memristor increases by $V_0\mathcal{T}_S(\mathcal{R}_M\mathcal{R}_S)$, the reflected voltage increases by $V_0\mathcal{T}_S(\mathcal{R}_M\mathcal{R}_S)\mathcal{R}_M$, while the transmitted voltage increases by $V_0\mathcal{T}_S(\mathcal{R}_M\mathcal{R}_S)\mathcal{T}_M$. After $N$-times back and forth bouncing, i.e. at time $\Delta t+N\cdot 2\Delta t$ the the incoming voltage increases by $V_0\mathcal{T}_S(\mathcal{R}_M\mathcal{R}_S)^N$, the reflected voltage increases by $V_0\mathcal{T}_S(\mathcal{R}_M\mathcal{R}_S)^N\mathcal{R}_M$, while the transmitted voltage increases by $V_0\mathcal{T}_S(\mathcal{R}_M\mathcal{R}_S)^N\mathcal{T}_M$. According to these considerations the incoming, reflected and transmitted voltage components can be added as a geometric series, i.e. right after $t=(2N-1)\Delta t$ the bias voltage on the memristor is calculated as:
\begin{equation} \label{kozep_mertaniosszeg}
       V_\mathrm{bias}((2N-1)\Delta t)= V_0 \mathcal{T}_S \cdot \frac{1-(\mathcal{R}_M \mathcal{R}_S)^N}{1 -(\mathcal{R}_M \mathcal{R}_S)} + V_0 \mathcal{T}_S \mathcal{R}_M\cdot \frac{1-(\mathcal{R}_M \mathcal{R}_S)^N}{1 -(\mathcal{R}_M \mathcal{R}_S)} -
    V_0 \mathcal{T}_S\mathcal{T}_M\cdot \frac{1-(\mathcal{R}_M \mathcal{R}_S)^N}{1 -(\mathcal{R}_M \mathcal{R}_S)}.
\end{equation}
From the $V_\mathrm{bias}((2N-1)\Delta t))=V_\mathrm{set}$ condition a voltage build-up time of 
\begin{equation}
        \tau_{0\rightarrow V\mathrm{set}}=   \Delta t\cdot \left[ \frac{\ln \left(1- V_\mathrm{set} \cdot \frac{1- \mathcal{R}_M\mathcal{R}_S }{ 2V_{0} \mathcal{T}_S \mathcal{R}_M} \right)}{\ln \left( \mathcal{R}_M\mathcal{R}_S)\right)}-1
        \right]
        \label{csunya_feszfelepules_logaritmusos_Tszamolasa}
\end{equation}                
is obtained.

\newpage

\section{Circuit-level modeling of resistance relaxation and set processes}\label{sec:simplemodel}

To incorporate resistance relaxation effects and finite set transition timescales of the VO$_2$ memristors into our circuit-level simulations, we constructed a custom memristor component in Matlab Simulink. In this case, the apporach is similar to our LTspice model consisting of a simple hysteretic switch controlled by the voltage drop on the memristor, i.e. a set (reset) transition occurs once the voltage grows above (decreases below) $V_\mathrm{set}$ ($V_\mathrm{reset}$). An important difference, however, that resistance change is not abrupt, rather it is governed by the following equations for its derivative, respectively describing the reset and the set transition:

\begin{equation}
    \frac{\text{d}R_M}{\text{d}t} =
\begin{dcases} 
\frac{R_\mathrm{M,OFF} - R_M}{\tau_{\rm reset}} & \\
\frac{R_\mathrm{M,ON} - R_M}{\tau_{\rm set}}. &  \\
   \end{dcases} 
\label{eq:simplemodel}
\end{equation}

\begin{figure}[b!]
    \centering
    \includegraphics[width=0.55\linewidth]{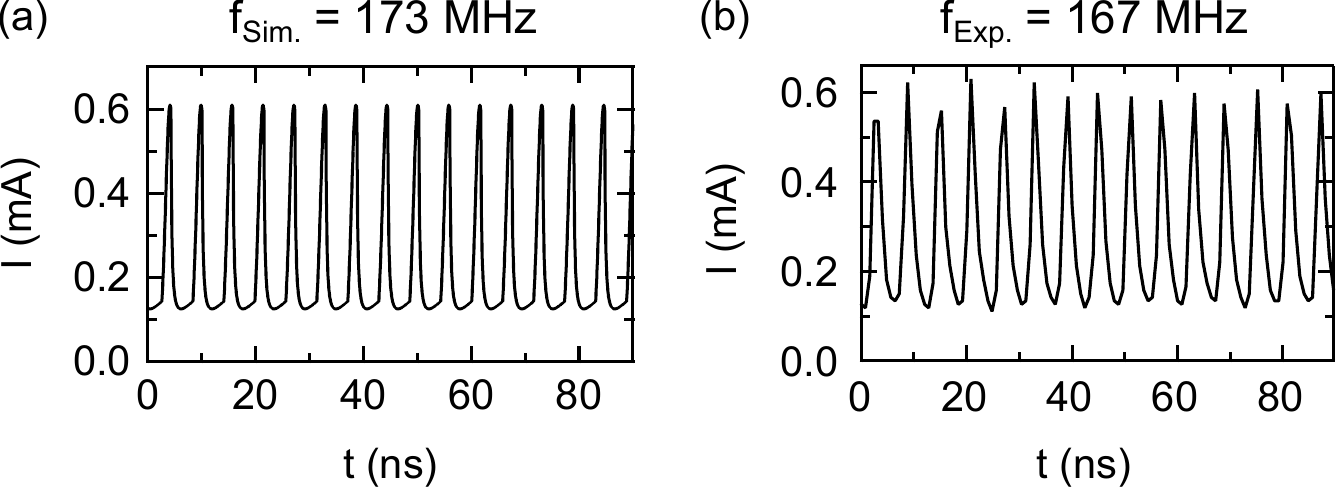}
    \caption{\it \textbf{Comparison of the simulated and experimental current waveforms.} (a) Simulated current trace exhibiting $f_{\rm Sim.}=173$~MHz oscillation frequency. The parameters of the simulated memristor are $V_\mathrm{set}=1.08$~V, $V_\mathrm{reset}=0.55$~V, $R_\mathrm{M,OFF}=8000~\Omega$, $R_\mathrm{M,ON} = 1000~\Omega$, and characteristic timescales of $\tau_{\rm set} = 600$~ps and $\tau_{\rm reset}=1.5$~ns. Driving voltage signal amplitude was set to $2V_0 = 8$~V and series resistor value of $R_S=30770~\Omega$ was used, along with $\Delta t = 32~$ps transmission line length (corresponding to $d = 6.5$~mm distance). (b) Experimental current trace of our highest frequency oscillation ($f_{\rm Exp.}=167$~MHz, taken from Fig. 1d of the main text).}
    \label{fig:sfig3}
\end{figure}

Here, $R_\mathrm{M,OFF}$ and $R_\mathrm{M,ON}$ characterize the resistance values towards the $R_M$ memristor resistance evolves over time, and the $\tau_{\rm set}$ and $\tau_{\rm reset}$ variables describe the timescales of the set process and the relaxation.
Fig.~\ref{fig:sfig3}a shows the resulting simulated current waveform of the memristor, where this simple model is applied with realistic memristor parameters. Here, $d=6.5~$mm series resistor-to-memristor distance was implemented, which is the same distance, as in the case of the fastest experimental oscillation depicted in Fig.~\ref{fig:sfig3}b. The oscillation frequency of the resulting simulated waveform yields $f_{\rm Sim.}=173~$MHz, assuming reasonable relaxation timescales of $\tau_{\rm set} = 600$~ps and $\tau_{\rm reset}=1.5$~ns. The simulation resembles the oscillation of the fastest experiment with $f_{\rm Exp.}=167~$MHz (Fig.~\ref{fig:sfig3}b, the same as Fig.~1d in the main text).

\newpage

\section{The role of delayed voltage build-up in stabilizing the oscillation}\label{sec:setresetanalysis}

In the limiting cases $d=0$ and $C=0$, i.e., when there is no delay in the voltage build-up, we can define the memristor resistances $R_\mathrm{set}$ and $R_\mathrm{reset}$, where the voltage on the memristor reaches the switching voltages $V_\mathrm{set}$ and $V_\mathrm{reset}$.
Reaching these resistances during oscillation is an essential condition for proper operation: once the set transition starts, the resistance should decrease enough to reach the reset voltage, and vice versa. In the $d=0$ and $C=0$ limit, however, the voltage drop on the memristor follows the resistance variation, i.e. when $R_\mathrm{reset}$ ($R_\mathrm{set}$) is reached, the memristor voltage already significantly goes below (above) the $V_\mathrm{set}$ ($V_\mathrm{reset}$) switching voltages. On the other hand, if the voltage goes significantly below the set voltage (above the reset voltage), we expect the switching to stop, i.e. we never reach the $R_\mathrm{reset}$ and ($R_\mathrm{set}$) resistances. Indeed, in a real VO$_2$ memristor, the bias voltage conditions affect switching dynamics to a great extent, i.e. the set (reset) process slows down, if the pulse amplitude (readout offset voltage) is not high (low) enough (see our pulsed measurements in Fig. 5b-g of the main text). To eliminate this problem, one can slow down the voltage variation by a finite distance $d$ or by a parallel capacitor $C$.

\begin{figure}[b!]
    \centering
    \includegraphics[width=1\linewidth]{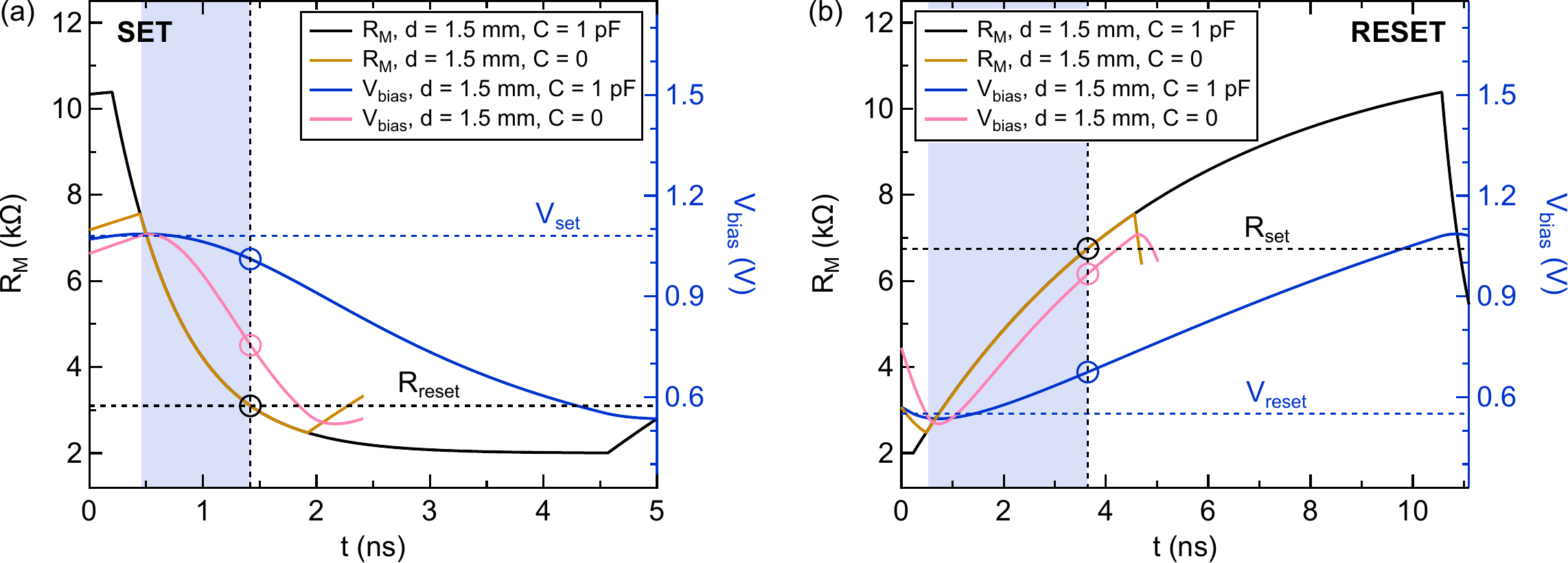}
    \caption{\it 
    \textbf{Illustrating the role of delayed voltage build-up in oscillation stabilization.} The parameters of the simulated memristor are $V_\mathrm{set}=1.08$~V, $V_\mathrm{reset}=0.55$~V (indicated by blue dashed lines), $R_\mathrm{M,OFF}=11600~\Omega$, $R_\mathrm{M,ON} = 2000~\Omega$, and characteristic timescales of $\tau_{\rm set} = 600$~ps and $\tau_{\rm reset}=5$~ns. Driving voltage signal amplitude was set to $2V_0 = 6$~V and series resistor value of $R_S=30770~\Omega$ was used, along with $\Delta t = 7.4~$ps transmission line length (corresponding to $d = 1.5$~mm distance). Two scenarios are compared: one without a capacitor, represented by the brown and pink traces corresponding to $R_M$ and $V_{\rm bias}$, and another where a $C = 1$~pF capacitor is connected in parallel with the memristor, shown by the black and blue traces for $R_M$ and $V_{\rm bias}$. (a) Comparison of the set processes and (b) reset processes are displayed, aligned by the $R_M$ traces in both cases. Horizontal black dashed lines indicate the $R_{\rm set} = 6750~\Omega$ and $R_{\rm reset}=3100~\Omega$ values which should be reached for switching, if the voltage drop on the memristor reaches $V_\mathrm{set}$ and $V_\mathrm{reset}$. The vertical black dashed lines emphasize the timesteps when $R_{\rm set}$ and $R_{\rm reset}$ are reached. The blue shaded areas highlight the region where the bias voltages diverge from each other in the two cases, before reaching $R_{\rm set}$ and $R_{\rm reset}$.}
    \label{fig:sfig2}
\end{figure}

In the following this phenomenon is illustrated by investigating the memristor resistance ($R_M$) and bias voltage ($V_{\rm bias}$) time traces obtained with the simple relaxation/set time model described in Section~\ref{sec:simplemodel}, concentrating closely on the set process (Fig.~\ref{fig:sfig2}a) and the reset process (Fig.~\ref{fig:sfig2}b). In these simulations, identical memristors were applied, with the same driving amplitudes and serial resistors and the same $d=1.5$~mm resistor-to-memristor distance. Two cases are compared, one without capacitor (brown and pink traces corresponding to $R_M$ and $V_{\rm bias}$) and one where $C=1$~pF capacitor was connected parallel to the memristor (black and blue traces corresponding to $R_M$ and $V_{\rm bias}$). 
Note, that in both cases, the rate of resistance change is solely governed by Equation~\ref{eq:simplemodel}, so the $R_M$ traces can be aligned with each other in the set and reset processes, as shown in Fig.~\ref{fig:sfig2}a and Fig.~\ref{fig:sfig2}b, respectively.
Naturally, in the real VO$_2$ memristors, the physical processes describing the set and reset processes are much more complicated than what is shown in idealized simulations with fixed $\tau_{\rm set}$ and $\tau_{\rm reset}$ time constants in Fig.~\ref{fig:sfig2}, yet, these simulations shed light on the connections between the internal switching timescales of the memristors and the characteristic timescales of the surrounding circuit.

In Fig.~\ref{fig:sfig2}a the set process is illustrated. At the beginning $V_\mathrm{set}$ is reached (blue dashed line), so $R_M$ starts to decrease. The goal is to reach $R_\mathrm{reset}$, i.e. the black dashed resistance value (see black circle). At finite $d=1.5\,$mm but no capacitance, the voltage variation is delayed compared to the resistance (pink curve), but the voltage would already decrease much below $V_\mathrm{set}$ when $R_\mathrm{reset}$ would be reached (see the pink encircled voltage level), i.e. the process would terminate before the next reset transition. If, however, the $C=1\,$pF capacitor is also applied, the voltage variation is even more delayed, and the corresponding blue voltage trace goes only a bit below $V_\mathrm{set}$ (see blue circle) until  $R_\mathrm{reset}$ is reached, i.e. the completion of the switching is reasonable. Obviously, the same result could be achieved without a parallel capacitor, but further increasing $d$.

Fig.~\ref{fig:sfig2}b illustrates the same for the reset process: after initially triggering the reset transition at $V_\mathrm{reset}$ (blue dashed line) the resistance should grow to $R_\mathrm{set}$ (black dashed line and circle). The delayed voltage variation of the blue curve aids this process by keeping the voltage close to $V_\mathrm{reset}$ until $R_\mathrm{set}$ is reached (see blue circle). Less delay (pink curve) would yield a significant voltage increase prior to reaching $R_\mathrm{set}$, which could terminate the switching.

\newpage
\section{Modeling bias voltage-dependent switching dynamics}

So far, our Matlab Simulink simulations considered constant (voltage-independent) time constants for the set and reset transitions (Eq.~\ref{eq:simplemodel}). We then extend this simplified model with a heuristic, voltage-dependent switching dynamics that mimics the fact in real memristors that the switching slows down and terminates as soon as the voltage goes significantly below (above) the set (reset) threshold voltage. Equation~\ref{eq:simplemodel} described in Section~\ref{sec:simplemodel} is extended by voltage-dependent terms as follows:

\begin{figure}[b!]
    \centering
    \includegraphics[width=0.7\linewidth]{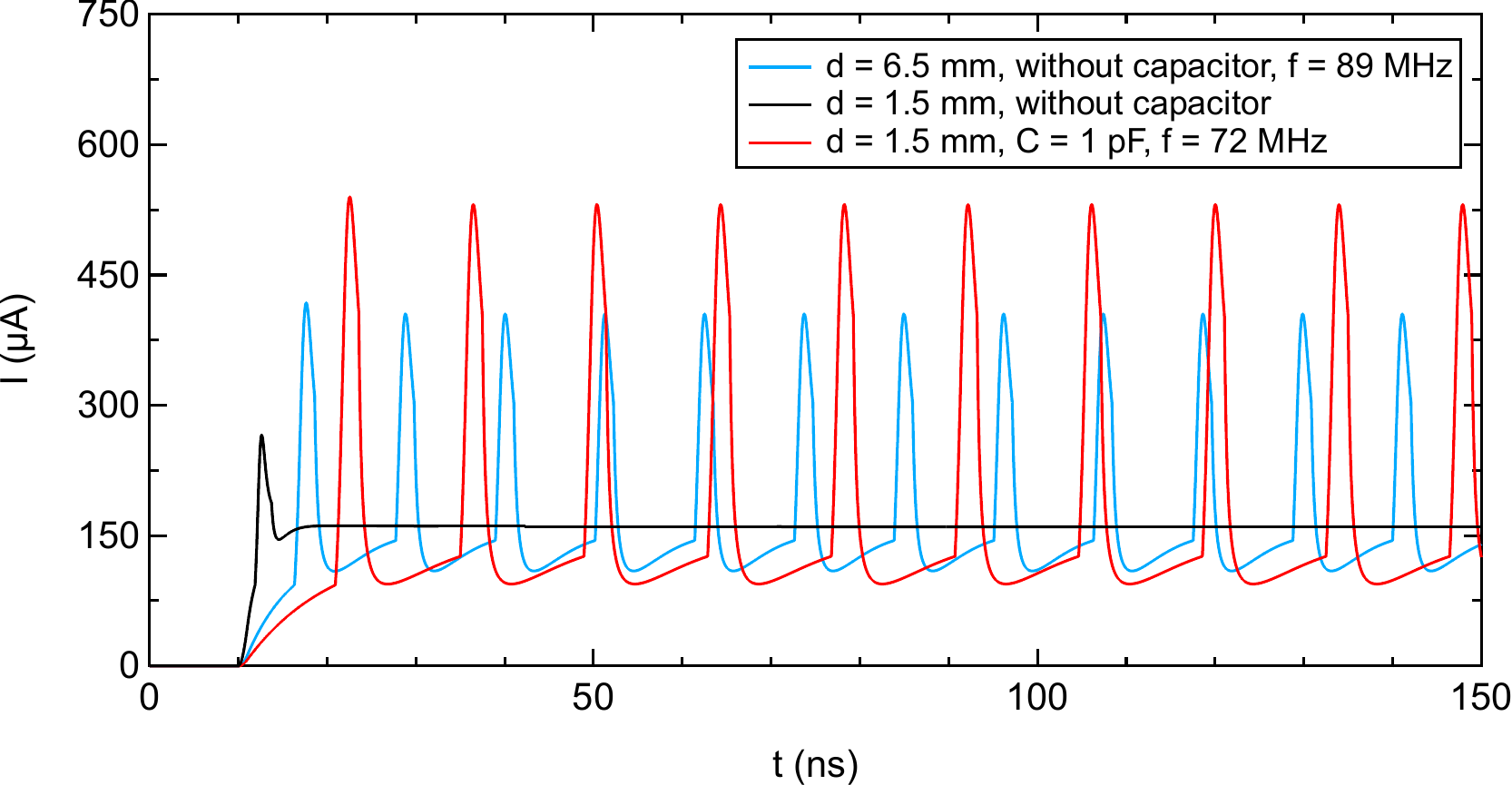}
    \caption{\it 
    \textbf{Analyzing the effect of bias voltage-dependent switching dynamics via simulation.} The parameters of the simulated memristor are $V_\mathrm{set}=1.08$~V, $V_\mathrm{reset}=0.55$~V, $R_\mathrm{M,OFF}=11600~\Omega$, $R_\mathrm{M,ON} = 100~\Omega$, and characteristic timescales of $\tau_{\rm set} = 600$~ps and $\tau_{\rm reset}=3.4$~ns. Driving voltage signal amplitude was set to $2V_0 = 6$~V and series resistor value of $R_S=30770~\Omega$ was used.
    Three scenarios are analyzed, one with $d=6.5$~mm ($\Delta t = 32~$ps), without capacitor (blue trace, $f=89~$MHz), one with $d=1.5$~mm ($\Delta t = 7.4~$ps)  without capacitor (black trace, no oscillation), and one with with $d=1.5$~mm ($\Delta t = 7.4~$ps), with $C=1~$pF (red trace, $f=72$~MHz). }
    \label{fig:sfig1}
\end{figure}

\begin{equation}
    \frac{\text{d}R_M}{\text{d}t} =
\begin{dcases} 
\frac{R_\mathrm{M,OFF} - R_M}{\tau_{\rm reset}} \cdot \left( \frac{V_{\rm set}-V_{\rm bias}}{\Delta V}\right)^2 & \text{(reset transition)} \\
\frac{R_\mathrm{M,ON} - R_M}{\tau_{\rm set}} \cdot \left( \frac{V_{\rm bias}-V_{\rm reset}}{\Delta V}\right)^2 & \text{(set transition)}. \\
   \end{dcases} 
\end{equation}\label{eq:Vdepmodel}

Here, $\Delta V =V_{\rm set}-V_{\rm reset}$, the voltage window where oscillation occurs. Note, that the above expressions realize a quadratic suppression of the derivative, as $V_{\rm bias} \rightarrow V_{\rm set}$ ($V_{\rm bias} \rightarrow V_{\rm reset}$), keeping the effective reset time (set time) close to $\tau_{\rm reset}$ ($\tau_{\rm set}$), when $V_{\rm bias} \approx V_{\rm reset}$ ($V_{\rm bias} \approx V_{\rm set}$) conditions persist.

Applying this model to a realistic device with our typical $d=6.5\,$mm ($\Delta t =32$~ps) transmission line length and no parallel capacitor, $f=89$~MHz oscillation frequency is retained, as shown by the blue current trace in Fig.~\ref{fig:sfig1}. However, when the transmission line length is decreased to $d=1.5$~mm (corresponding to $\Delta t =7.4$~ps), oscillations disappear (see black trace in Fig.~\ref{fig:sfig1}), which aligns well with our experimental observation reported in Figure 4e of the main text, where on-chip resistors with $d=1.5$~mm effective distance were applied. By the addition of $C=1$~pF parallel capacitor while keeping $d=1.5$~mm, oscillation is recovered, resulting in $f=72~$MHz frequency, similarly to our experiments.

The message of simulations presented here and in Section~\ref{sec:setresetanalysis} is, that for reaching stable oscillations, it is essential to tune the timescales of the surrounding circuit such that it is slightly slower than the internal set- and relaxation timescales of our VO$_2$ memristor. This can be achieved by either including a parallel capacitor or a larger distance between the memristor and the serial resistor, which ensures the appropriate voltage conditions on the device to reach large enough resistance changes to realize stable oscillation.

\bibliographystyle{achemso}
\bibliography{References}